\documentclass[%
reprint,
superscriptaddress,
 amsmath,amssymb,
aps,
prb,
]{revtex4-2}

\usepackage{graphicx}
\usepackage{dcolumn}
\usepackage{bm}
\usepackage{hyperref}
\usepackage{cleveref}
\hypersetup{
    colorlinks = true,
    allcolors=black,
    citecolor=blue,
    linkcolor=blue
} 
\usepackage{amsmath, amsfonts, dsfont, mathrsfs, float,amssymb}
\usepackage{color}
\usepackage{subcaption}

\newcommand{\barj}[1]{\bar{\jmath}}

\begin{document}


\title{Identifying Many-Body Localization in Realistic Dot Arrays}

\author{Alexander Nico-Katz}
\affiliation{Department of Physics and Astronomy, University College London, London WC1E 6BT, United Kingdom}

\author{Gulzat Jaliel}
\affiliation{London Centre for Nanotechnology, University College London, 17-19 Gordon Street, London WC1H 0AH, United Kingdom}

\author{Madhumita Sarkar}
\affiliation{Department of Physics and Astronomy, University College London, London WC1E 6BT, United Kingdom}

\author{Paola Atkinson}
\affiliation{Cavendish Laboratory, University of Cambridge, J. J. Thomson Avenue, Cambridge CB3 0HE, United Kingdom}
\affiliation{Sorbonne Université, CNRS, Institut des NanoSciences de Paris, INSP, 4 place Jussieu, F-75005 Paris, France}

\author{Thomas A. Mitchell}
\affiliation{Cavendish Laboratory, University of Cambridge, J. J. Thomson Avenue, Cambridge CB3 0HE, United Kingdom}

\author{David A. Ritchie}
\affiliation{Cavendish Laboratory, University of Cambridge, J. J. Thomson Avenue, Cambridge CB3 0HE, United Kingdom}

\author{Charles G. Smith}
\affiliation{Cavendish Laboratory, University of Cambridge, J. J. Thomson Avenue, Cambridge CB3 0HE, United Kingdom}

\author{Sougato Bose}
\affiliation{Department of Physics and Astronomy, University College London, London WC1E 6BT, United Kingdom}

\date{\today}
             
\begin{abstract}
    We investigate the feasibility of identifying many-body localization in semiconductor quantum dot arrays within current experimental capabilities. Using an extended Fermi–Hubbard model parameterized by realistic device conditions, we analyze the phase diagram through a set of observables with varying experimental accessibility. Our results demonstrate that signatures of MBL can, in principle, be accessed in present-day quantum dot platforms. Crucially, we show that the widely used imbalance fails to distinguish between interaction-induced freezing and genuine localization, highlighting the importance of carefully selecting and interpreting experimental probes. We identify observables that can reliably distinguish genuine MBL from interaction-induced effects, thereby establishing practical criteria for detecting localization in realistic quantum dot arrays.
\end{abstract}

\maketitle
\section{Introduction}
\label{sec:intro}
Many-body localization (MBL), the breakdown of ergodicity and the suppression of transport in disordered interacting quantum systems \cite{Nandkishore2015, Alet2018, Abanin2017,Abanin2019}, has emerged as a central paradigm in non-equilibrium quantum physics. Unlike conventional thermalizing systems, MBL systems fail to act as their own heat bath and therefore do not satisfy the eigenstate thermalization hypothesis \cite{Deutsch1991, Srednicki1994, Nandkishore2015, Abanin2019}. As a consequence, they retain memory of their initial conditions for arbitrarily long times, exhibit suppressed transport, and display a characteristic logarithmic growth of entanglement despite the absence of thermalization \cite{Bardarson2012, Serbyn2013, Huse2014, Abanin2019, PhysRevB.105.024301}. Experimental signatures of MBL have been observed in a variety of platforms, including ultracold atoms and trapped ions in optical lattices \cite{Schreiber2015, Luschen2017, Choi2016, Smith2016}, as well as superconducting qubit systems \cite{Roushan2017}, with recent studies suggesting that transmon-based quantum processors may operate near the boundary between localized and chaotic regimes \cite{Berke2022}.

A natural and largely unexplored platform for realizing MBL is provided by semiconductor quantum dot arrays (QDAs). These systems offer highly tunable realizations of fermionic lattice models in one and two dimensions \cite{Hensgens2017, mukhopadhyay2018}, with flexible control over geometry and interactions. As such, they present an appealing setting in which to investigate localization phenomena in regimes complementary to existing platforms. However, current-generation devices are limited in size and are subject to noise and fabrication constraints, posing challenges for both accessing and identifying MBL.

In this work, we investigate the emergence of MBL in realistic QDA systems by combining experimentally motivated modeling with numerical analysis. Starting from the characterization of a state-of-the-art double-dot device, we extract physically relevant parameter regimes for an extended Fermi–Hubbard model describing larger arrays. We then map out the phase diagram of this model across these parameter ranges, identifying regimes consistent with ergodic, localized, and intermediate behavior.
\begin{figure}[ht]
    \centering
    \includegraphics[width=\linewidth]{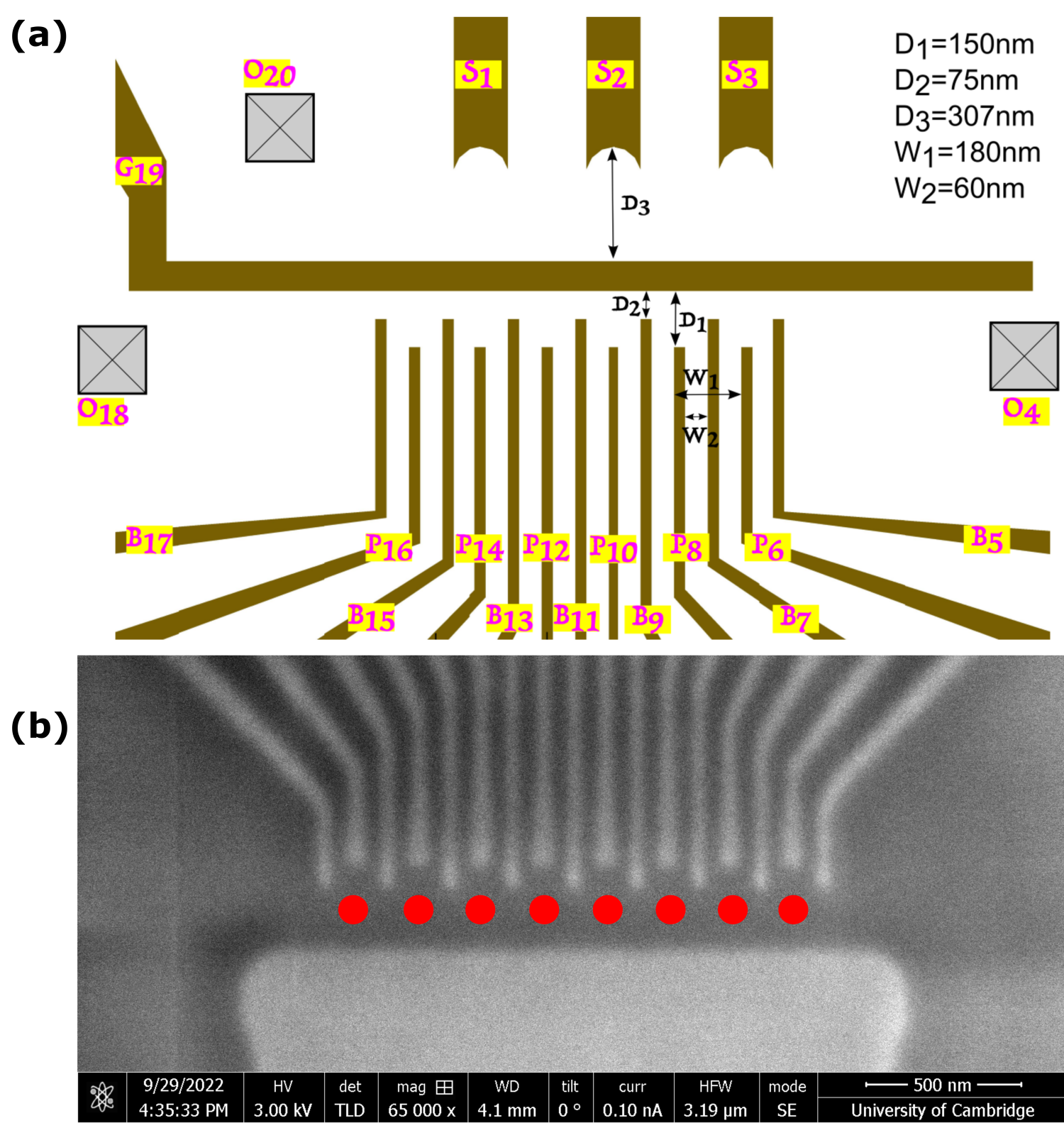}
    \caption{\textbf{(a)} The six-dot design for the specific  experimental device we characterize. In \cref{sec:experiment} we extract rough parameter ranges for the theoretical model that the device simulates from the pair of dots defined by the gates $G_{19}-B_{15}-B_{13}-B_{11}$ and addressed by plunger gates $P_{14}$ and $P_{12}$ respectively. \textbf{(b)} An SEM image of an array we fabricated which is similar to the design we characterize. Slight blurring of the SEM image is due to a layer of protective PMMA with an approximate thickness of 200 nm.}
    \label{fig:schematic}
\end{figure}

To characterize these phases, we analyze a set of diagnostic observables in both global and local forms, ranging from quantities requiring partial state tomography to experimentally accessible local charge measurements. By comparing their behavior across the phase diagram, we establish a hierarchy of probes that can reliably distinguish MBL from ergodic dynamics in finite-sized systems. Our results provide both a theoretical characterization of localization in experimentally relevant parameter regimes and a practical protocol for detecting MBL in quantum dot arrays using minimal measurement overhead.

In \cref{sec:model} we introduce and discuss the model that quantum dot arrays simulate. In \cref{sec:sigs} we analyze the various phases of the model and define several quantities in both bulk and local variants to differentiate MBL from the other phases of the model. We discuss our experimental devices, namely one-dimensional lateral arrays of electrostatically defined quantum dots, and then characterize such a device to extract rough ranges for the theoretical model parameters from experimental data in \cref{sec:experiment}. Finally, in \cref{sec:phases} we develop a protocol for identifying MBL with minimal measurements on a realistic device.
\begin{figure}[ht]
    \centering
    \includegraphics[width=\linewidth]{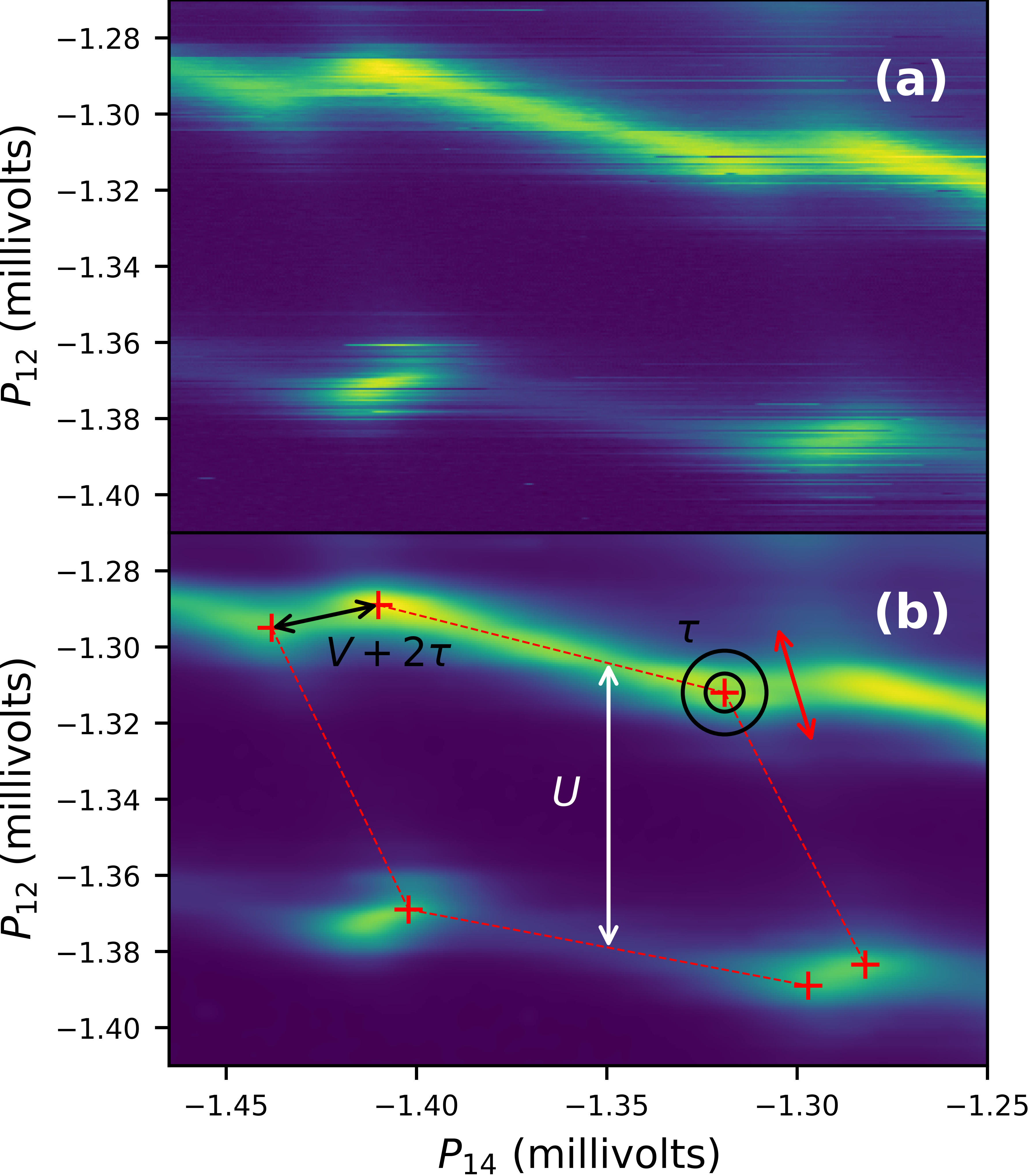}
    \caption{\textbf{(a)} A typical experimental honeycomb cell of the charge-stability diagram obtained by differential conductance measurements on the two dots defined by the gates $G_{19}-B_{15}-B_{13}-B_{11}$ in the middle of a device similar to that shown in \cref{fig:schematic}. Bright (dark) regions indicate higher (lower) measured values of the differential conductance as a function of the two local plunger gate voltages $P_{14}$ and $P_{12}$. \textbf{(b)} A Gaussian smoothing of the raw data shown in panel \textbf{(a)} with a schematic overlay of how the parameters of the theoretical model that the device simulates are extracted from the geometry of the honeycomb cell.}
    \label{fig:experimental}
\end{figure}

\section{Model Hamiltonian}
\label{sec:model}

We consider the extended Fermi--Hubbard model described by the Hamiltonian
\begin{equation}\label{eqn:fermihubbard}
H = \tau \sum_j^{L-1} \left(c^\dagger_{j} c_{j+1} + \text{h.c.}\right) + V \sum_j^{L-1} n_j n_{j+1} + \sum_j^L h_j n_j,
\end{equation}
where $c_j$ ($c_j^\dagger$) are fermionic annihilation (creation) operators at site $j$, and $n_j = c_j^\dagger c_j$. The parameters $\tau$ and $V$ denote the tunnelling and nearest-neighbour Coulomb interaction energies, respectively, while $h_j$ are random on-site energies drawn uniformly from $[-h, h]$, with $h$ setting the disorder strength. We assume (i) a single spin species and (ii) a single active energy level per site. This effective spinless description can be realized by lifting the spin degeneracy with an applied magnetic field. We further assume that the electrostatically determined model parameters, namely the onsite charging energy, nearest neighbour interaction strength, and tunnelling amplitudes, are not significantly modified by the magnetic field within the experimentally relevant regime where orbital effects remain negligible. This assumption is commonly adopted in electrostatically defined Fermi Hubbard models~\cite{Hensgens2017, HensgensThesis}.

Using a Jordan--Wigner transformation, the Hamiltonian in \cref{eqn:fermihubbard} maps onto an extended XXZ spin chain that exhibits multiple competing phases making it a suitable model to study whether MBL can be experimentally distinguished from other phases. A conducting phase for $0 < V/\tau < 2$, an interaction-induced insulating phase for $V/\tau > 2$, and a many-body localized phase at sufficiently large disorder $h/\tau$. In the non-interacting limit $V/\tau = 0$, the system reduces to a disordered free-fermion model exhibiting Anderson localization for any $h/\tau > 0$ in the thermodynamic limit. In our work we focus on the regime $0 < V/\tau < 2$, where the ergodic-MBL transition is well defined, and aim to determine whether MBL can be reliably distinguished from other ergodicity-breaking mechanisms in realistic experimental settings. 

Recent work indicates that reliable characterization of the ergodic-MBL transition requires access to exponential time and length scales in microscopic analyses~\cite{Panda2019, Sierant2022}, and that finite-size transitions may belong to a different universality class than their thermodynamic-limit counterparts~\cite{Khemani2017universality}. Rather than attempting to resolve critical behavior, we instead focus on identifying distinct regimes and determining the conditions under which they can be \textit{differentiated} in realistic experimental quantum dot arrays.

\section{Experimental System and Parameter Extraction}
\label{sec:experiment}

The experimental devices we consider are lateral arrays of electrostatically defined quantum dots, formed by selectively depleting electrons using nano-fabricated gate electrodes on the surface of a GaAs/AlGaAs heterostructure. As a concrete realization of this architecture, the benchmark device consists of a linear array of six quantum dots defined by the gate pattern shown in \cref{fig:schematic}\textbf{(a)}. A scanning electron microscope (SEM) image of a similar eight-dot device is shown in \cref{fig:schematic}\textbf{(b)}, with the dot positions indicated by red circles. These devices are scaled-up versions of standard multi-dot architectures~\cite{Braakman2013, Hensgens2017}. The device is fabricated on a Si-doped $\text{GaAs}/\text{Al}_x\text{Ga}_{1-x}\text{As}$ heterostructure hosting a two-dimensional electron gas $90\,\text{nm}$ below the surface, with mobility $9\times10^5\,\text{cm}^2/\text{Vs}$ and electron density $1.62\times10^{11}\,\text{cm}^{-2}$. Measurements are performed in a dilution refrigerator at $T\sim 70\,\text{mK}$, with an electron temperature of $\sim 100\,\text{mK}$. Coulomb diamond measurements yield an on-site charging energy $E_C \sim 1.3\,\text{meV}$ and level spacing $\sim 400\,\mu\text{eV}$. Charge-qubit coherence times are taken to be $1$--$10\,\text{ns}$~\cite{Hayashi2003, Petersson2010}, while silicon-based and spin-qubit devices exhibit significantly longer coherence times~\cite{Gorman2005, Petta2005, Bluhm2011, Kawakami2016}.

Tunnelling between adjacent dots is controlled via the barrier gates $B_j$, while on-site chemical potentials are tuned by the plunger gates $P_j$, with the outermost dots coupled to electronic reservoirs. The central gate $G_{19}$ (\cref{fig:schematic}\textbf{(a)}) defines a global barrier, and the sensing gates $S_{1(2)(3)}$ act as quantum point contact charge detectors, whose non-linear conductance probes the local electrostatic environment~\cite{Field1993}. Capacitive cross-talk between gates is mitigated using virtual gates, i.e., linear combinations of physical gate voltages that independently control electrochemical potentials or tunnel barriers~\cite{Hensgens2017,Volk_2019}.

Next we focus our attention to the extraction of the parameter ranges (for $U/\tau$ and $V/\tau$) of our theoretical model \cref{eqn:fermihubbard} from the experimental data shown in \cref{fig:experimental}\textbf{(a)}. This informs the model regimes that the experimental device can access.

To extract model parameters, we probe the central double-dot defined by $G_{19}-B_{15}-B_{13}-B_{11}$ via differential conductance measurements while sweeping $P_{12}$ and $P_{14}$, yielding a honeycomb charge-stability diagram in which high-conductance regions mark degeneracies between charge configurations. A representative raw cell is shown in \cref{fig:experimental}\textbf{(a)}. To facilitate parameter extraction, the data is processed using a Gaussian kernel-density estimate, producing the smoothed honeycomb cell shown in \cref{fig:experimental}\textbf{(b)}, from which the parameters of the extended Fermi-Hubbard model can be determined~\cite{VanDerWiel2002, HensgensThesis}.


The on-site interaction $U$ is extracted as the energy cost of adding an electron to a single dot, corresponding to the vertical separation between adjacent charge configurations in the honeycomb diagram~\cite{VanDerWiel2002, Hensgens2017}. In \cref{fig:experimental}, we select a cell without contributions from zero-dimensional level spacing, yielding a conservative estimate where $U/\tau$ is not enhanced by such effects; nevertheless, $U/\tau$ remains sufficiently large to justify assumption (ii) in \cref{sec:model}.

The nearest-neighbour interaction $V$ is inferred from the separation between charge configurations differing by one additional electron on both sites. Due to hybridization induced by $\tau$, this separation scales as $V+2\tau$~\cite{HensgensThesis}, and thus requires an independent estimate of $\tau$.

\begin{figure*}
    \begin{subfigure}[b]{0.3\textwidth}
          \centering
          \includegraphics[width=\linewidth]{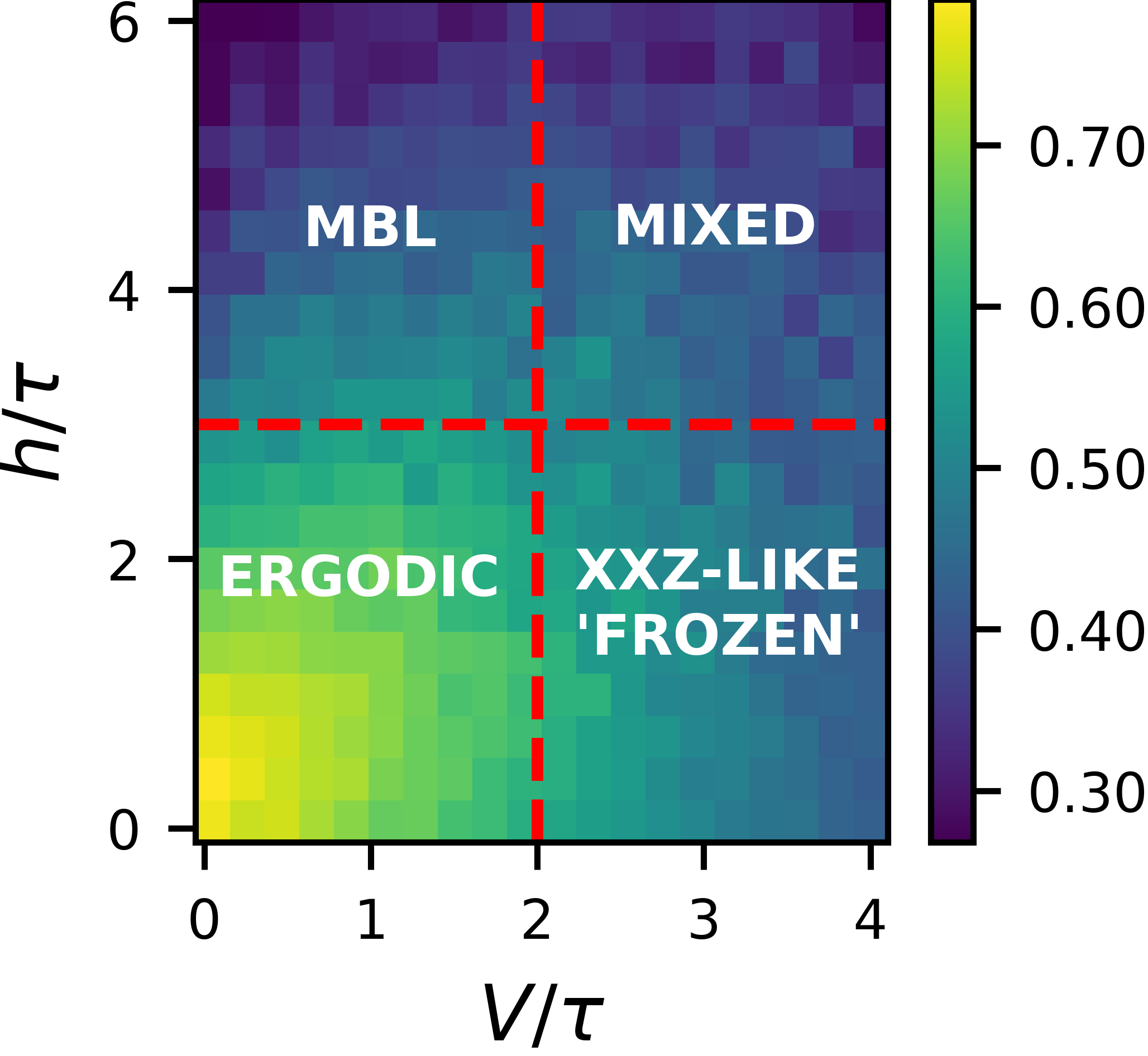}
          \caption{Bulk von Neumann Entropy}
     \end{subfigure}
     \begin{subfigure}[b]{0.3\textwidth}
          \centering
          \includegraphics[width=\linewidth]{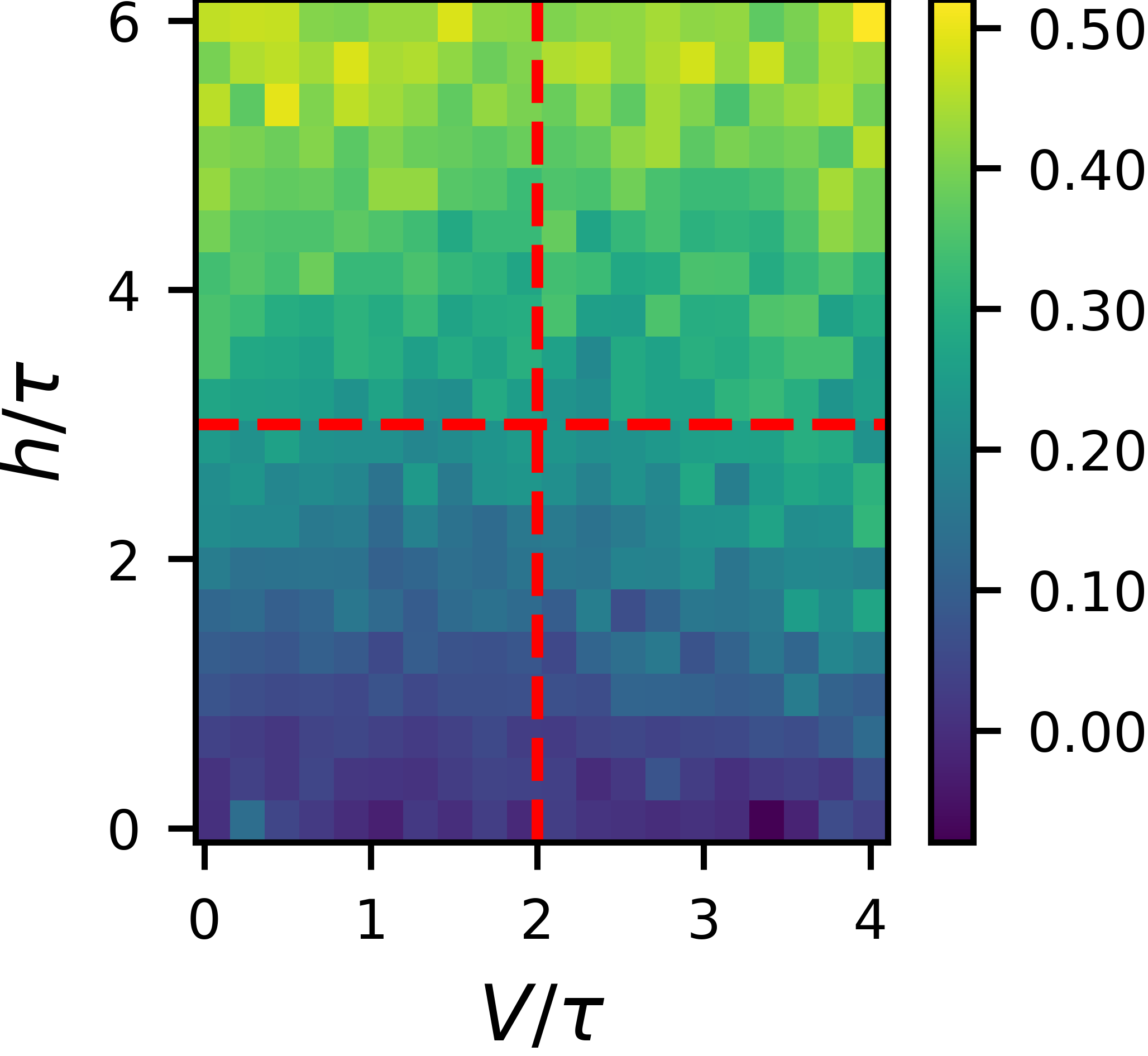}
          \caption{Bulk Imbalance}
     \end{subfigure}
     \begin{subfigure}[b]{0.3\textwidth}
          \centering
          \includegraphics[width=\linewidth]{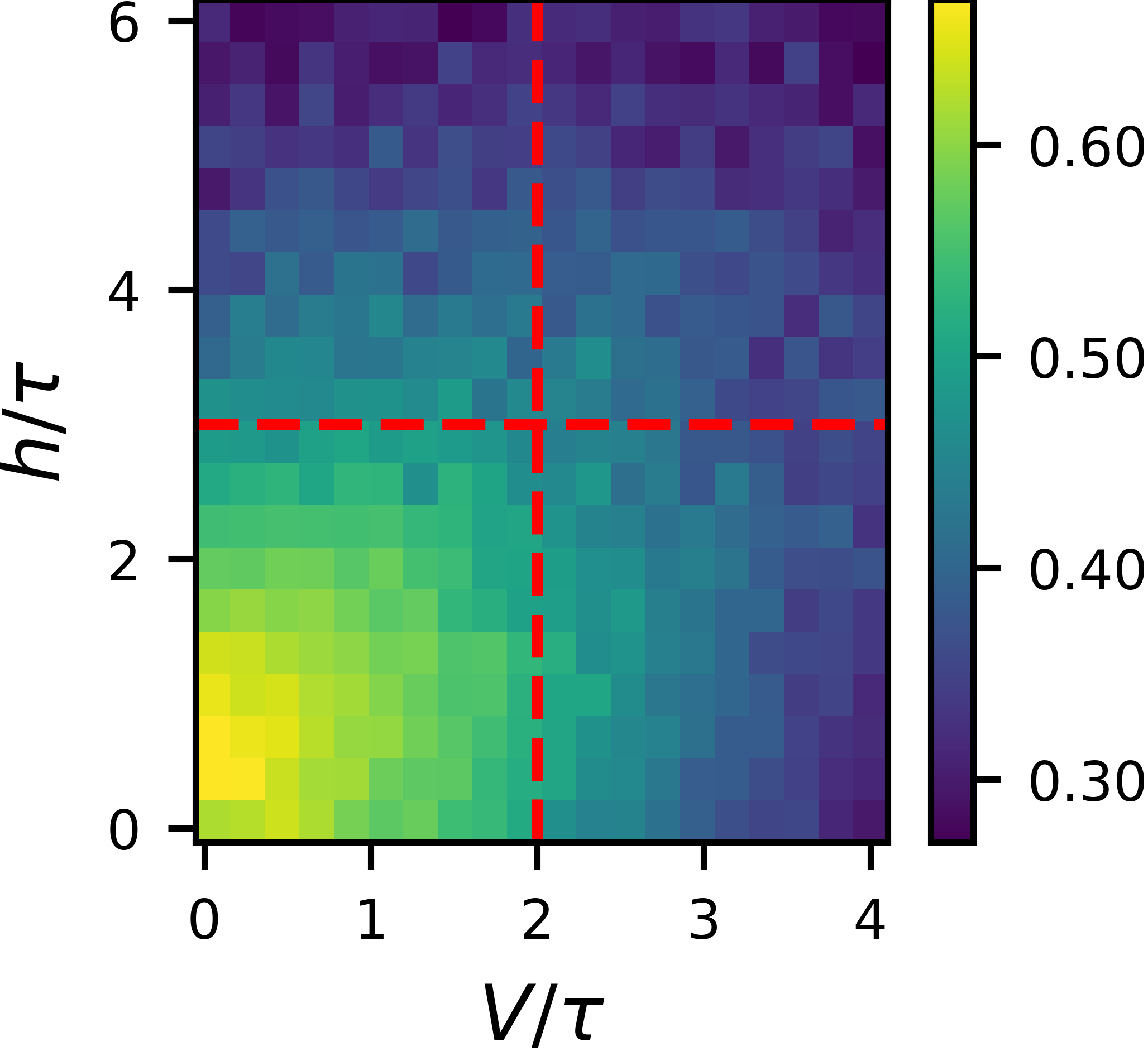}
          \caption{Bulk Number Entropy}
     \end{subfigure}
     
     \begin{subfigure}[b]{0.3\textwidth}
          \centering
          \includegraphics[width=\linewidth]{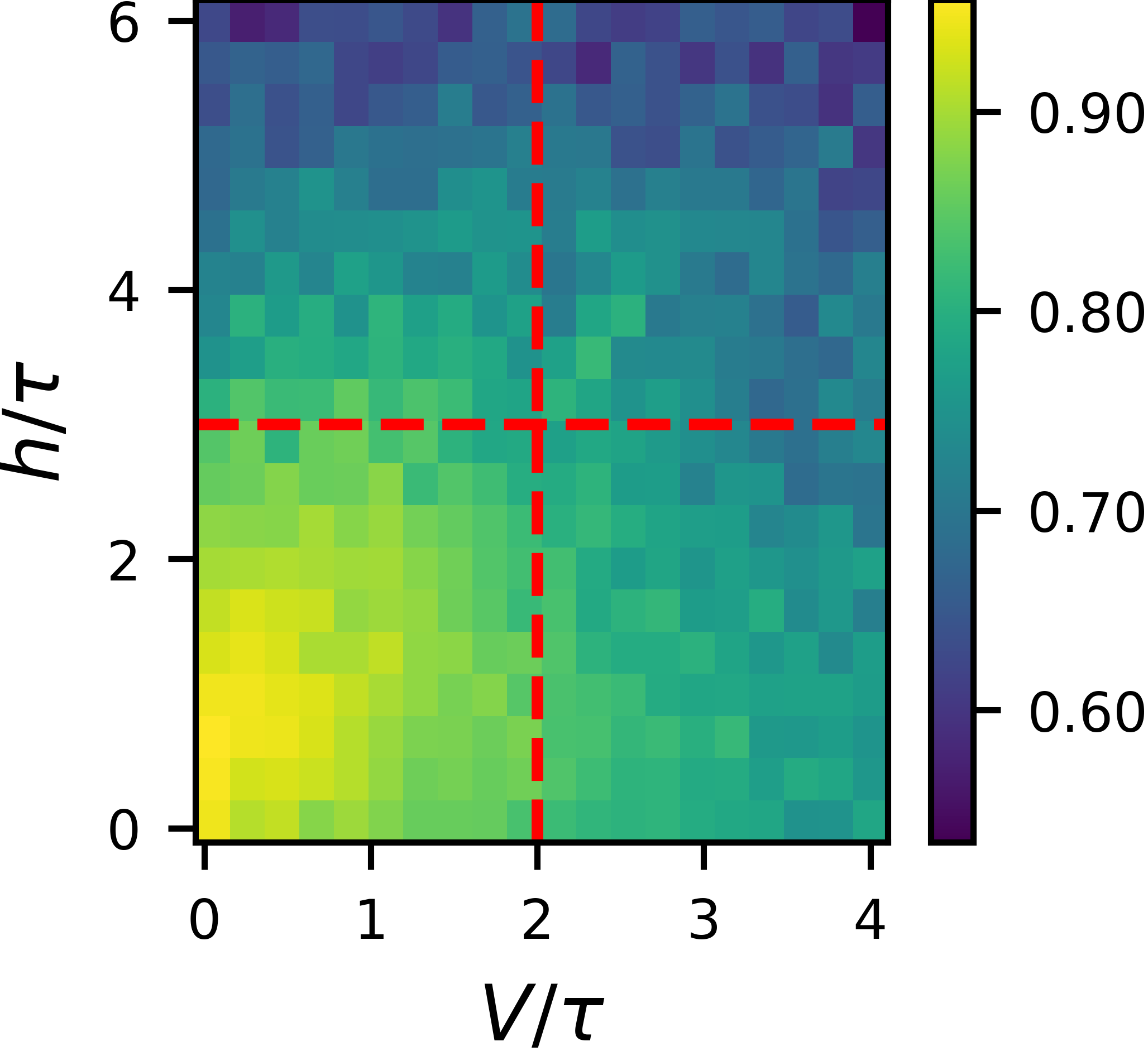}
          \caption{Local von Neumann Entropy}
     \end{subfigure}
     \begin{subfigure}[b]{0.3\textwidth}
          \centering
          \includegraphics[width=\linewidth]{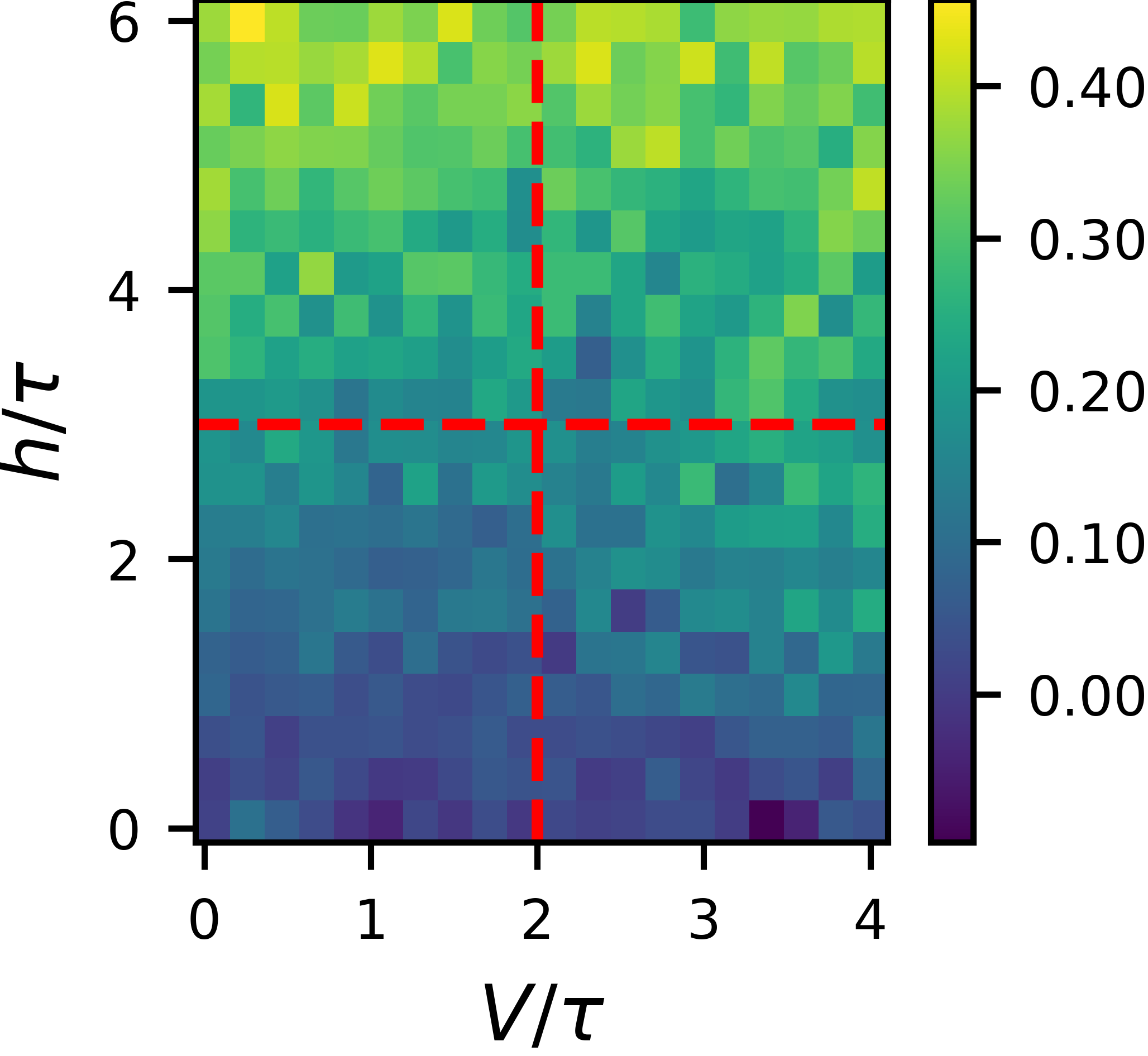}
          \caption{Local Imbalance}
     \end{subfigure}
     \begin{subfigure}[b]{0.3\textwidth}
          \centering
          \includegraphics[width=\linewidth]{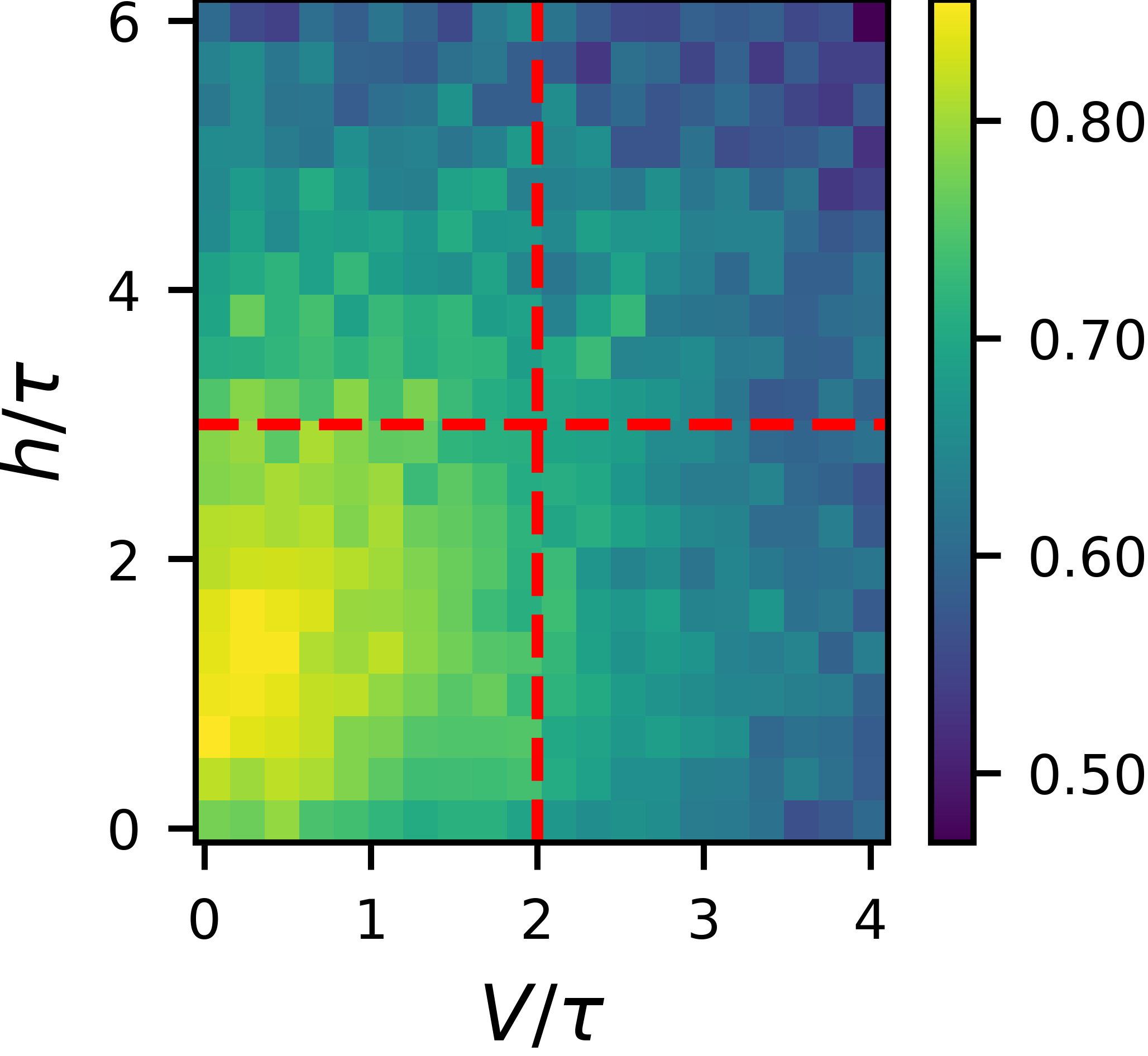}
          \caption{Local Number Entropy}
     \end{subfigure}
     
    \caption{Disorder-averaged late-time quantities across the full $V-h$ phase diagram for the realistic system; initialized in a charge-density wave configuration. The realistic system consists of $L=8$ sites, total evolution times of $\tau t_f = 150$, no time-averaging (the final value at $t_f$ is simply read out), and disorder-averaging over only $50$ samples per $(V, h)$ coordinate.}
    \label{fig:phase}
\end{figure*}

The tunnelling amplitude $\tau$ manifests through hybridization of neighbouring charge states, leading to rounding of phase boundaries and broadening of triple points in the stability diagram. It can be extracted via several methods, including analysis of boundary curvature~\cite{Wang2011}, time-domain oscillations~\cite{Shi2013, Wang2017}, and fitting conductance line cuts~\cite{VanDerWiel2002, Hensgens2017, mukhopadhyay2018}. In practice, we estimate $\tau$ by identifying the minimum and maximum radii of smeared triple points in \cref{fig:phase}, which also provide bounds on $V+2\tau$.

Altogether, the geometry of the charge-stability diagram determines ranges of $U/\tau$ and $V/\tau$, as illustrated in \cref{fig:experimental}\textbf{(b)}. By identifying triple points (black crosses) and their broadening (black circles), we extract bounds on $\tau$, $V+2\tau$, and $U$, summarized in \cref{tab:table1}. In all cases, $U \gg \{\tau, V\}$, validating the single-level approximation and consistent with previous multi-dot experiments~\cite{Hensgens2017}. Finally, the disorder strength $h/\tau$ is independently tunable via plunger gate voltages, subject only to the constraint $U \gg h$.

\begin{table}[ht]
    \centering
    \renewcommand{\arraystretch}{1.5}
    \begin{tabular}{|c||c|c|} \hline
         Parameters & Minimum $\tau$ & Maximum $\tau$ \\ \hline\hline
         $V/\tau$ & 3.727 & 0.603 \\
         $U/\tau$ & 12.941 & 4.792 \\ \hline
    \end{tabular}
    \caption{Table of upper and lower bounds on the considered parameter ranges for \cref{eqn:fermihubbard} extrapolated from the features of the experimental charge stability diagram of \cref{fig:experimental}. }
    \label{tab:table1}
\end{table}

\section{Probing Quantities}
\label{sec:sigs}

We consider three quantities: the von Neumann entropy $S$, the widely-used imbalance $\mathcal{I}$, and the number entropy $S_N$. These differ both in their experimental accessibility and, as we show in \cref{sec:phases}, in their ability to differentiate regimes in the phase diagram of \cref{eqn:fermihubbard}.

The first quantity of interest is the von Neumann entropy which, in a bipartite pure state, unambiguously quantifies the entanglement across the bi-partition. It has been used extensively in the context of MBL theory and experiment, and serves here as a benchmark for the other quantities~\cite{Bardarson2012, Serbyn2013, Xu2018, Lukin2019, Yousefjani2022}. For a given subsystem $\rho(t)$ of size $L_\rho$ it is defined as 
\begin{equation}\label{eq:vne}
    S\left(\rho(t)\right) = -\frac{1}{L_\rho}\text{Tr}\left[\rho(t) \log_2 \rho(t) \right]
\end{equation}
where we have chosen a logarithm base 2 such that the resulting quantity is measured in bits, and that it saturates (for a single species of fermion with local dimension 2, and given that $L_\rho / L \leq 1/2$) to unity. In experiment, the calculation of \cref{eq:vne} would require full tomography of the density operator of the region of interest which is a prohibitively expensive and difficult task, but it serves as a crucial benchmark nonetheless.

\begin{table}[ht]
    \centering
    \renewcommand{\arraystretch}{1.5}
    \begin{tabular}{|c||c|c|} \hline
         Quantity & Bulk & Local \\ \hline\hline
         VN Entropy & $S^{(b)}(t) = S(\rho_\text{hc}(t))$ & $S^{(l)} = S(\rho_\text{2}(t))$ \\
         Imbalance & $\mathcal{I}^{(b)}(t) = \mathcal{I}(\rho_\text{f}(t))$ &
        $\mathcal{I}^{(l)}(t) = \mathcal{I}(\rho_\text{2}(t))$ \\
         Number Entropy & $S_N^{(b)}(t) = S_N(\rho_\text{hc}(t))$ & $S_N^{(l)}(t) = S_N(\rho_\text{2}(t))$ \\
         \hline
    \end{tabular}
    \caption{ Summary of bulk and local variants of the quantities discussed in \cref{sec:sigs}. Bulk quantities are calculated over the state of the full system $\rho_\text{f}(t)$ or half the chain $\rho_\text{hc}(t)$, local quantities are calculated over two sites in the middle of the device $\rho_2(t)$. Experimentally, the imbalance and number entropy require identical local charge-sensing measurements only, and highly experimentally tractable. }
    \label{tab:table2}
\end{table}
The second quantity we consider is the imbalance, widely used in MBL~\cite{Schreiber2015, Luitz2016, Bordia2017, Luschen2017b, NicoKatz2022}. The imbalance quantifies the deviation of the system from its initial charge configuration and is directly related to standard auto-correlation functions. We define it as
\begin{equation}\label{eq:imb}
    \mathcal{I}(\rho(t)) = \frac{2}{L_\rho}\sum_{j} \text{Tr}\left[\rho(0) n_j \right]\text{Tr}\left[\rho(t) n_j \right]
\end{equation}
where $j$ runs over the physical sites of the subsystem $\rho(t)$. In the case of an initial charge density wave state, for which $n_j$ has $\rho(0)$ as an eigenstate, this expression reduces to the conventional definition of the imbalance as the difference between occupancy numbers on odd and even sites. If the charge configuration of $|\psi(t)\rangle$ becomes uncorrelated with the initial configuration of $|\psi(0)\rangle$, the imbalance decays to zero, $\mathcal{I}(t) = 0$, whereas persistent (anti-)correlations result in a finite non-zero value, $\mathcal{I}(t) > 0$ ($\mathcal{I}(t) < 0$). In experiment, $\mathcal{I}(\rho(t))$ requires only charge sensing on the relevant sites, which is significantly easier than the full state tomography required to compute the von Neumann entropy.

Finally we consider the number entropy, a quantity that has been used in studies of MBL~\cite{Lukin2019, Emmanouilidis2020, Ghosh2022}, and which is simply the entropy of the discrete probability distribution $p(\rho(t), n)$ of finding $n$ particles in the subsystem $\rho(t)$, defined as
\begin{equation}\label{eq:nent}
    S_N(t) = - \frac{1}{L_\rho} \sum_\mathcal{N} p(\rho(t), n=\mathcal{N}) \log_2 p(\rho(t), n=\mathcal{N}).
\end{equation}
This quantity is directly related to the von Neumann entropy $S$ by $S = S_N + S_C$ where $S_C$ is the configurational entropy: the contribution to the entanglement due to configurational correlations. We can compute the distributions from the density operator by constructing projectors $P_{\mathcal{N}} = \sum_r |\mathcal{N}_r\rangle\langle \mathcal{N}_r|$ where $|\mathcal{N}_r\rangle$ are the $\mathcal{N}$-particle states in the number basis that span the reduced Hilbert space of $\rho(t)$. The probability is then given by $p(\rho(t), n=\mathcal{N}) = \text{Tr}[\rho(t) P_{\mathcal{N}}]$. We calculate an ergodic limit of the number entropy in \cref{sec:app-thermalentropy}, which we use to benchmark our numerical results.

We consider both bulk and local variants of the quantities by evaluating three different subsystems: the full system $\rho_\text{f}(t) = |\psi (t) \rangle\langle  \psi (t)|$, half of the system $\rho_\text{hc}(t)$, and two sites in the middle of the system $\rho_\text{2}(t)$. The quantities are computed using the equations in \cref{sec:sigs} using the bulk or local reduced density operators according to \cref{tab:table2}.

In experiment the number entropy requires the same charge sensing measurements as the imbalance, but yields signatures similar to the von Neumann entropy. This becomes useful in \cref{sec:phases} where we find that the number entropy can identify phases that the imbalance cannot without requiring the prohibitively difficult state tomography of the full von Neumann entropy. Moreover, the local variants of the quantities shown in the last column of \cref{tab:table2} require measurements only on two sites, making their calculation even simpler in experiment.

\begin{figure*}
    \begin{subfigure}[b]{0.45\textwidth}
          \centering
          \includegraphics[width=\linewidth]{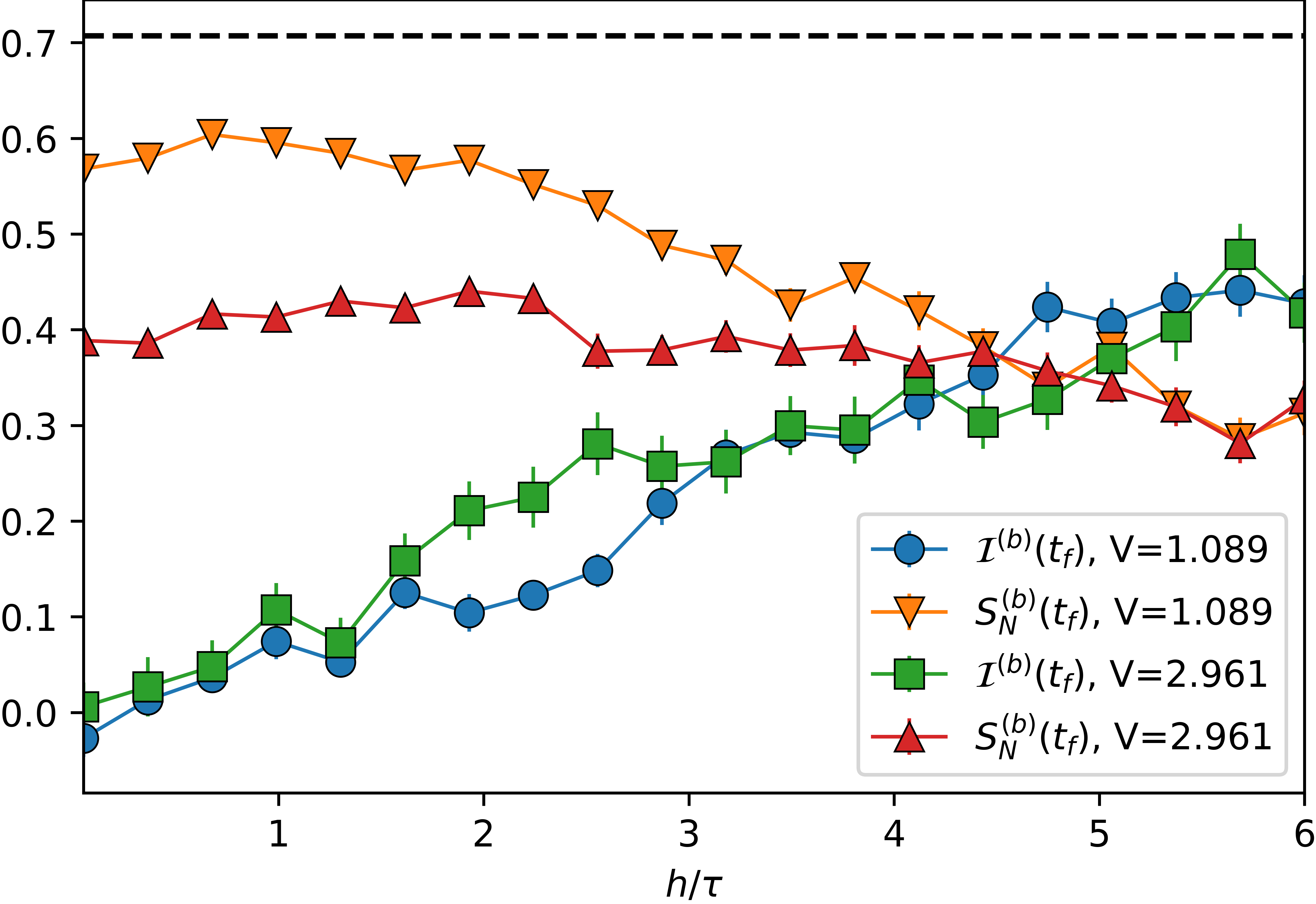}
          \caption{Bulk Imbalance and Number Entropy against $h/\tau$}
     \end{subfigure}
     \begin{subfigure}[b]{0.45\textwidth}
          \centering
          \includegraphics[width=\linewidth]{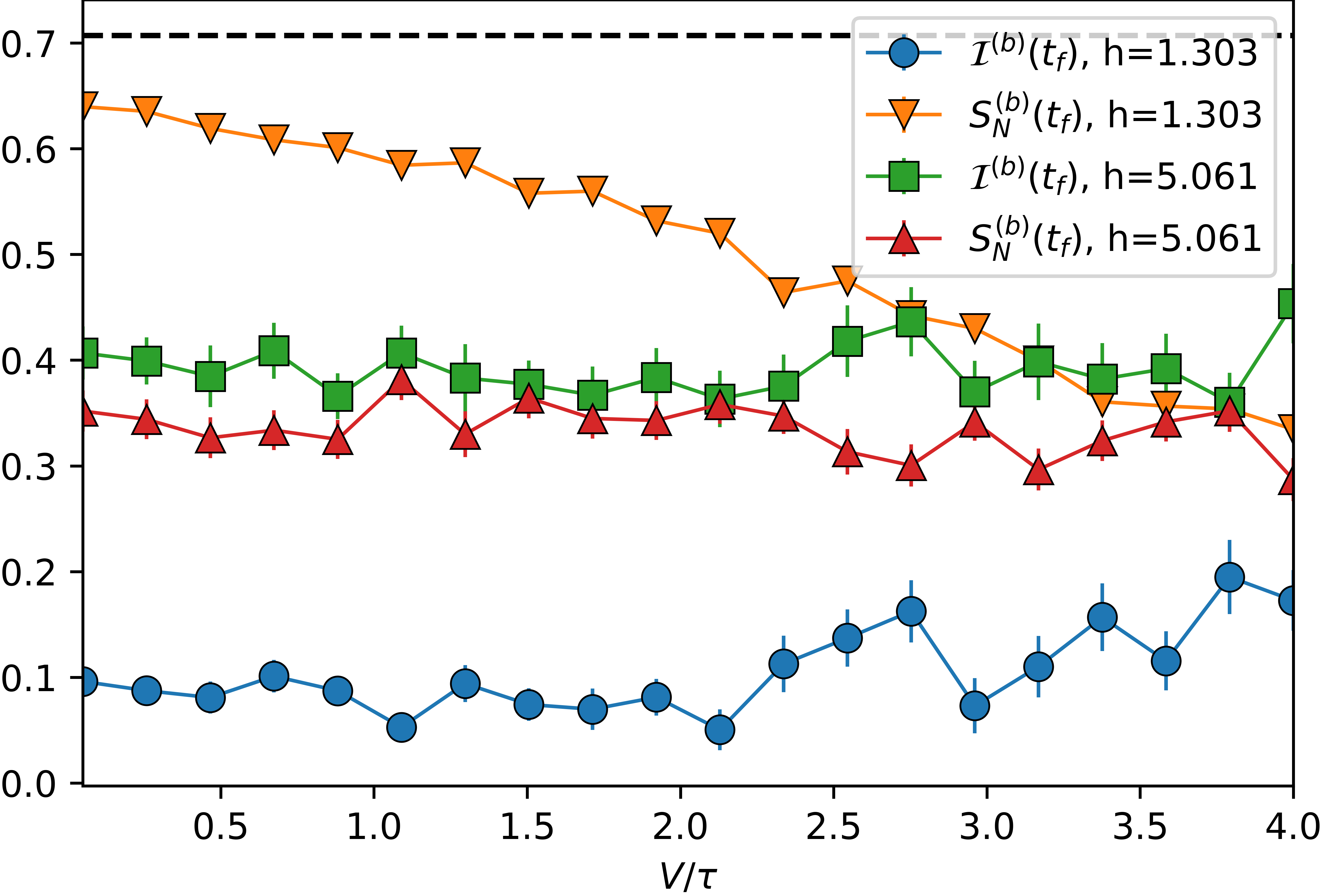}
          \caption{Bulk Imbalance and Number Entropy against $V/\tau$}
     \end{subfigure}
     
     \begin{subfigure}[b]{0.45\textwidth}
          \centering
          \includegraphics[width=\linewidth]{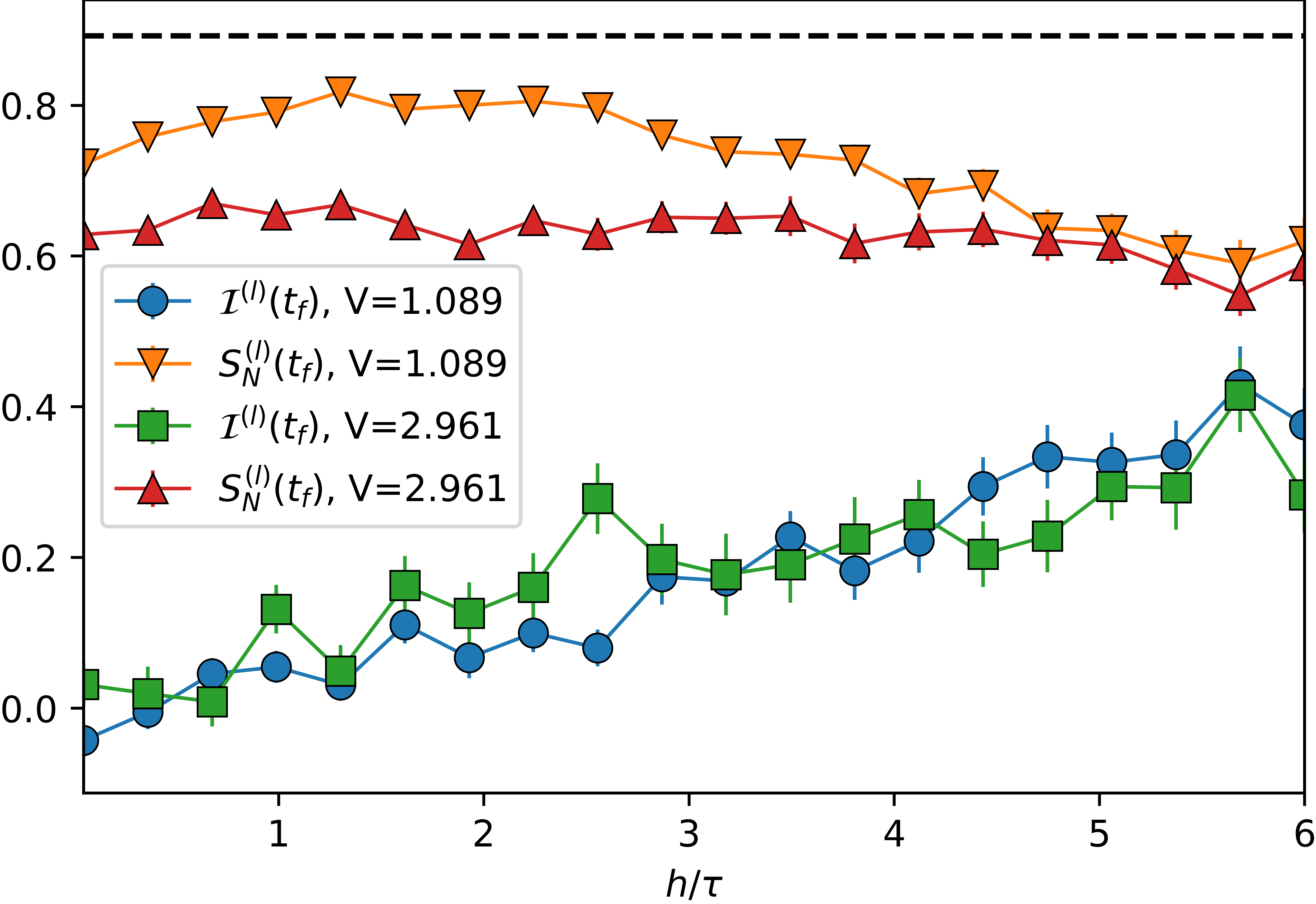}
          \caption{Local Imbalance and Number Entropy against $h/\tau$}
     \end{subfigure}
     \begin{subfigure}[b]{0.45\textwidth}
          \centering
          \includegraphics[width=\linewidth]{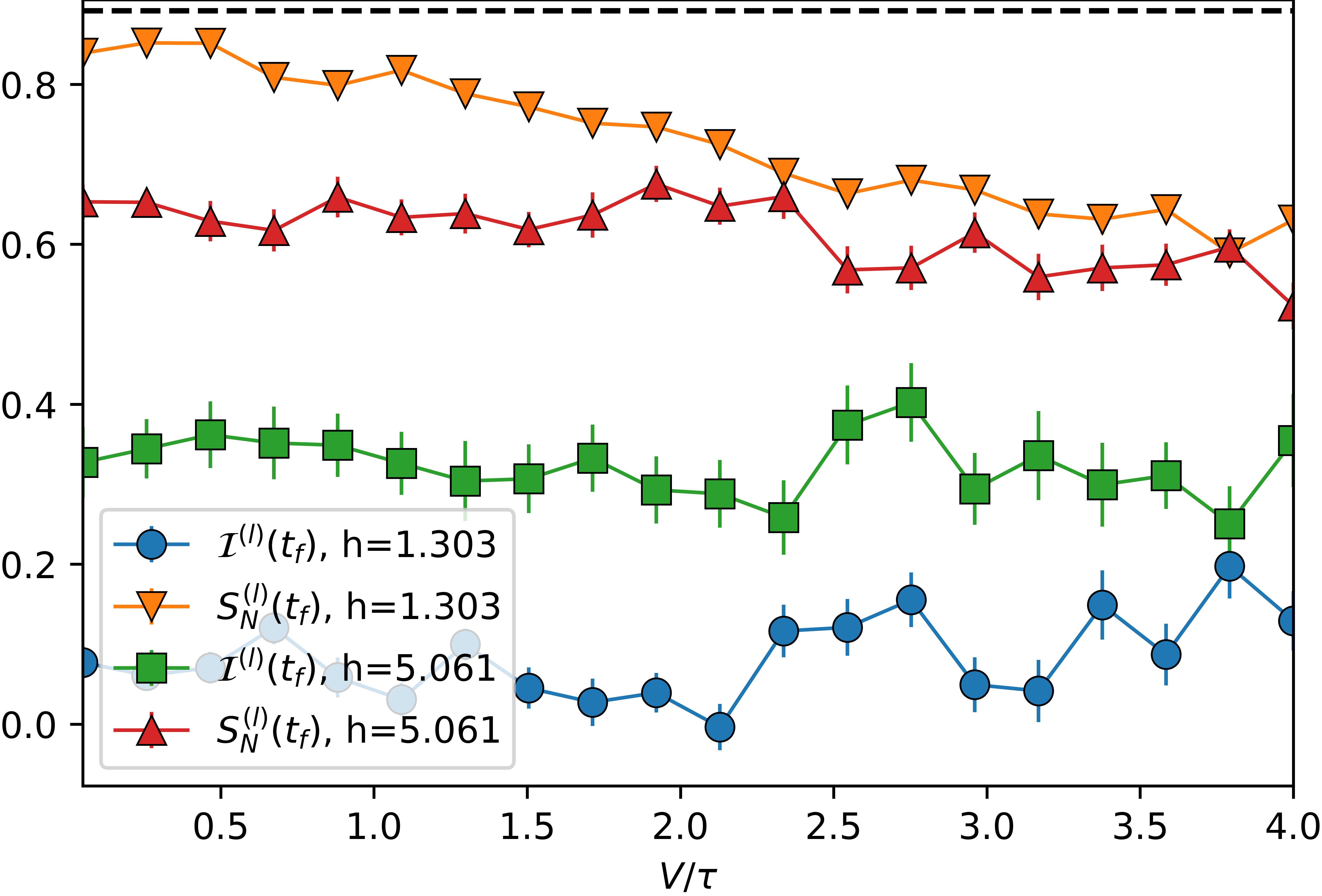}
          \caption{Local Imbalance and Number Entropy against $V/\tau$}
     \end{subfigure}
     \caption{ Slices of the phase diagrams of \cref{fig:phase} (the realistic $L=8$ model) for \textbf{(a)} and \textbf{(c)} fixed values of $V/\tau$ on both sides of the transition point between thermal and insulating phases at $V/\tau=2$, and \textbf{(b)} and \textbf{(d)} fixed values of $h/\tau$ on both sides of the crossover from thermal to MBL regimes at  $h/\tau=3.5$; error bars shown where visible. }
     \label{fig:phase_slice}
\end{figure*}

\section{Signatures of Localization in Realistic Dot Arrays}
\label{sec:phases}
We investigate the phase diagram of \cref{eqn:fermihubbard} at a final time $t_\text{f}$ following initialization in a charge density wave state, where every other site is occupied at $t=0$. 

To this end, we vary $V/\tau$ and $h/\tau$, and evaluate the bulk and local observables defined in \cref{sec:sigs}. Guided by the experimentally accessible parameter ranges extracted in \cref{sec:experiment}, we consider $V/\tau \in [0,4]$ and $h/\tau \in [0,6]$. 
To model realistic devices, we restrict to system size $L=8$, evolution times $\tau t_\text{f} = 150$ (corresponding to $t_\text{f} \sim 1$--$10\,\text{ns}$ for $\tau \sim 50$--$100\,\mu\text{eV}$, within typical coherence times~\cite{Hayashi2003, Petersson2010}), no time-averaging, and $50$ disorder realizations. We additionally include absolute ($\pm 0.1$) and relative ($\pm 5\%$) box-distributed uncertainties in $V/\tau$ and $h/\tau$ to account for cross-talk and fabrication imperfections. Results for an idealized system are presented in \cref{sec:app-phases-ideal}. 

We remark that our choice of measurement times $t_f$ are comparable with modern characterizations of readout times \cite{bohuslavskyi2022fast}, and an actual experimental implementation may smear our results. We argue against this issue in two ways: firstly, and most importantly, the observables considered in this work depend only on charge occupation probabilities, and not on phase coherence. Thus one can simply pinch-off the sites being measured by quickly raising the barrier gates, a process which takes place on nanosecond timescales (c.f. Ref.~\cite{wenz2019quantum} and supplementary material thereof). This fixes the charge number which can then be read out over longer timescales than those determined by the many-body dynamics of the system. Secondly it is the \textit{relative} strength of $h/\tau$ and $V/\tau$ which determines the overall timescale of dynamics in the system. $V/\tau$ is fixed, but the actual relative strength of disorder to $\tau$ depends on the individual realization. Thus different realizations of different disorder landscapes will slightly modify the timescale of dynamics; and hence the observable smearing issue raised in this comment is partially taken into account during disorder-averaging. This is a subtle and minor point regarding implicit averaging, and we defer a full characterization of the interplay between time and disorder-averaging to future research.

Secondly, and more compellingly, observables considered in this work depend only on charge occupation probabilities measured via charge sensing, and not on phase coherence. Thus one can simply pinch-off the sites being measured by quickly raising the barrier gates (on the order of nanoseconds, e.g. \cite{wenz2019quantum} and supplementary material thereof). This fixes the charge number which can then be read out over longer timescales than those determined by the many-body dynamics of the system.

The resulting phase diagrams are shown in \cref{fig:phase}, where panels \textbf{(a)}--\textbf{(c)} display the bulk von Neumann entropy, imbalance, and number entropy, and panels \textbf{(d)}--\textbf{(f)} their local counterparts. Red dashed lines indicate the expected transition at $V/\tau = 2$ and an ergodic--MBL crossover at $h/\tau \sim 3$. The von Neumann entropy and number entropy exhibit similar qualitative behaviour, with large values in the thermal regime that decrease as either $h/\tau$ or $V/\tau$ increases.

In contrast, the imbalance remains largely insensitive to the transition at $V/\tau = 2$, staying close to its thermal value for sufficiently low disorder. This indicates that the imbalance alone cannot distinguish between the MBL regime and an interaction-induced insulating regime: a transition observed as a function of $h/\tau$ may occur within a phase that is already non-ergodic due to strong interactions. The number entropy resolves this ambiguity. While requiring the same charge-sensing measurements as the imbalance, it is sensitive to both the MBL transition and the interaction-driven transition at $V/\tau = 2$. Thus, by post-processing the same measurement data in two different ways, one can distinguish between thermal, MBL, and insulating regimes.

A key observation is that the qualitative features of the bulk and local phase diagrams are consistent. This suggests that local measurements on only a few sites suffice to identify the relevant regimes. Combined with the above, this leads to a simple experimental protocol: perform charge-sensing measurements on a small subset of dots, and compute both the local imbalance and number entropy. Upon scanning $h/\tau$, a transition in both quantities indicates MBL, whereas a transition only in the imbalance signals an interaction-induced insulating regime.

We attribute the insensitivity of the imbalance at large $V/\tau$ to a coherent “rolling” dynamics of the charge-density wave. In this regime, strong interactions suppress neighbouring excitations, leading to oscillations between configurations such as $|\circ, \bullet, \circ \cdots \rangle$ and $|\bullet, \circ, \bullet \cdots \rangle$, consistent with blockade-like physics in related systems~\cite{Lin2019, Bernien2017, Khemani2019}. This yields a thermal-like imbalance while producing a sharply peaked particle-number distribution and hence a low number entropy. We support this interpretation with numerical analysis in \cref{sec:app-oscillating}, and observe similar behaviour in systems with odd system sizes.

To further mimic experimental protocols, we consider one-dimensional scans of the phase diagram in \cref{fig:phase_slice}, where either $V/\tau$ or $h/\tau$ is varied while the other is held fixed. Panels \textbf{(a)} and \textbf{(b)} show bulk results, and panels \textbf{(c)} and \textbf{(d)} show local results. The dashed lines indicate the thermal value of the number entropy derived in \cref{sec:app-thermalentropy}. As expected, the imbalance is largely insensitive to $V/\tau$, while the number entropy depends strongly on both $V/\tau$ and $h/\tau$, clearly distinguishing regimes when one parameter is varied at low values of the other. However, local measurements alone may be insufficient when scanning only $h/\tau$, suggesting that either bulk observables or scans in $V/\tau$ may be required.

Finally, we note that our use of limited disorder averaging and large parameter uncertainties represents a conservative scenario; experimental data are likely to be clearer. Results for ideal systems (\cref{sec:app-phases-ideal}) confirm that distinguishing MBL from interaction-induced insulation becomes significantly easier under improved conditions. Nevertheless, even in this realistic setting, local charge sensing suffices to identify MBL and distinguish it from other ergodicity-breaking mechanisms in current-generation quantum dot arrays.

\section{Discussions}
\label{sec:conc}

Our results demonstrate that MBL is, in principle, accessible in state-of-the-art quantum dot arrays, but that its unambiguous identification is non-trivial. A central challenge lies in the interplay between disorder and strong interactions: while large on-site charging energies are required to suppress double occupancy, sufficiently strong disorder must also be applied to induce localization without exceeding these energy scales. Although reducing the tunnelling amplitude $\tau$ can enhance localization, this places more stringent demands on coherence times, as longer evolution times are required to observe the relevant dynamics. Improvements in device design, such as enhanced isolation from phonon environments, may help mitigate these constraints.

Within these limitations, we constructed a realistic model based on experimental device parameters and analyzed its phase diagram as a function of disorder and interaction strength. We identified an interaction-induced insulating regime that can mimic signatures of MBL, highlighting that distinguishing localization mechanisms is essential in experimental settings. In particular, we find that the widely-used imbalance is largely insensitive to this regime and therefore insufficient on its own to identify MBL.

We instead propose a protocol based on the number entropy, which requires the same charge-sensing measurements as the imbalance but is sensitive to both disorder-driven and interaction-driven transitions. By comparing imbalance and number entropy extracted from the same measurements, one can reliably distinguish thermal, MBL, and interaction-induced insulating regimes. Importantly, we show that local measurements on only a few sites yield qualitatively similar information to bulk observables, significantly reducing experimental overhead.

Taken together, our results establish practical criteria for identifying MBL in realistic quantum dot arrays, and demonstrate that careful selection of experimentally accessible observables is crucial for distinguishing genuine localization from other ergodicity-breaking mechanisms in current-generation devices.

\section{Acknowledgements}
\label{sec:ack}

The authors acknowledge the EPSRC grant Nonergodic quantum manipulation EP/R029075/1. MS and SB also acknowledge EPSRC-SFI funded project EP/X039889/1 (GeQuantumBus). The authors also thank Dr. G. Stefanou for his advice and technical experimental support.

\section{Author Contributions}
\label{sec:contribs}

A.N.-K., G.J., C.G.S., and S.B. devised the project and the main conceptual ideas. A.N.-K. developed the theoretical model and carried out the numerical simulations. G.J. designed, planned, and carried out the experiments. P.A., T.A.M., and D.A.R. contributed to the fabrication of the devices. A.N.-K., M.S. and G.J. wrote the manuscript with support and contributions from C.G.S. and S.B. M.S. substantially rewrote and restructured the manuscript and led the preparation of the referee responses during peer review.

\appendix

\section{Mapping the Fermi-Hubbard Model to the XXZ Spin Chain}
\label{sec:app-jordanwigner}
The spinless Fermi-Hubbard model of \cref{eqn:fermihubbard} can be mapped onto an spin-1/2 XXZ model by means of a standard Jordan-Wigner transformation. We start by identifying the two states of each site - occupied or unoccupied - with the two spin-1/2 states - spin-up and spin-down. We then relate the fermionic creation and annihilation operators with the spin-1/2 raising and lowering operators $c^\dagger_j \to S^+_j$ and $c_j \to S^-_j$, and impose the standard fermionic anti-commutators by applying highly non-local Jordan-Wigner string operators such that the final transformation takes the form:
\begin{align}\label{eqn:jwcreate}
    c_j^\dagger = \left(\prod_{k < j} \sigma^z_j\right)S^+_j \\ \label{eqn:jwdestroy}
    c_j = \left(\prod_{k < j} \sigma^z_j\right)S^-_j
\end{align}
where the $\sigma^z_j$ is the standard Pauli Z operator. In practice, due to the fact that $(\sigma^z_j)^2 = 1$, many of these Jordan-Wigner strings cancel and one can often manipulate the final Hamiltonian into a local form. Substituting \cref{eqn:jwcreate} and \cref{eqn:jwdestroy} into the Hamiltonian \cref{eqn:fermihubbard}, and simplifying the Jordan-Wigner strings, yields the model
\begin{align*}
    H = \tau &\sum_j^{L-1} \left(S_j^+ \sigma_j^z S_{j+1}^- + \text{h.c.}\right)\\ 
    &+ V \sum_j^{L-1} S_j^+ S_j^- S_{j+1}^+ S_{j+1}^- + \sum_j^L h_j S_j^+ S_j^-
\end{align*}
which we can simplify further by noting that $S_j^+ S_j^- = S^z_j + 1/2$, that $S_j^+ \sigma_j^z = - S_j^+$, and by invoking the definition of $S^\pm_j = S_j^x \pm i S^y_j$; where $S_j^\alpha$ are the standard spin-1/2 operators. Under these simplifications, the model now takes the form
\begin{align*}
    H = - 2 \tau &\sum_j^{L-1} \left(S_j^x S_{j+1}^x + S_j^y S_{j+1}^y \right) \\ 
    &+ V \sum_j^{L-1} \left(S_j^z S_{j+1}^z + S_j^z + S_{j+1}^z + 1\right) \\ 
    &+ \sum_j^L h_j \left(S_{j}^z + 1/2\right)
\end{align*}
wherein the negative overall phase of the tunnelling term proportional to $t$ is irrelevant to the physics of the system: we can e.g. simply relabel every other site such that, for odd $j$, $S^{x/y}_j \to - S_j
^{x/y}$, and $S_j^z$ is left unchanged. We eliminate all constant terms which do not contribute to the physics of the system, and note that - since the Hamiltonian conserved total particle number, its Jordan-Wignerization equivalently conserves total spin. Thus we can neglect all terms proportional to $\sum_j^L S_j^z$, which appear in the sum over single-site operators proportional to $V$ such that $\sum_j^{L-1} S_j^z + S_{j+1}^z  = -S_1^z - S_L^z + \sum_j^L S_j^z $. Collecting terms, and performing these substitutions gives us an extended XXZ Hamiltonian
\begin{align*}
    H = 2\tau &\sum_j^{L-1} \left(S_j^x S_{j+1}^x + S_j^y S_{j+1}^y + \frac{V}{2 t} S_j^z S_{j+1}^z \right) \\
    & - V\left(S_1^z + S_L^z\right) + \sum_j^L h_j S_{j}^z
\end{align*}
which exhibits an XXZ transition at $V = 2\tau$, makes clear the importance of edge effects in small systems by explicit inclusion of the dangling $S_1^z$ and $S_L^z$ operators.

\begin{figure*}
    \begin{subfigure}[b]{0.3\textwidth}
          \centering
          \includegraphics[width=\linewidth]{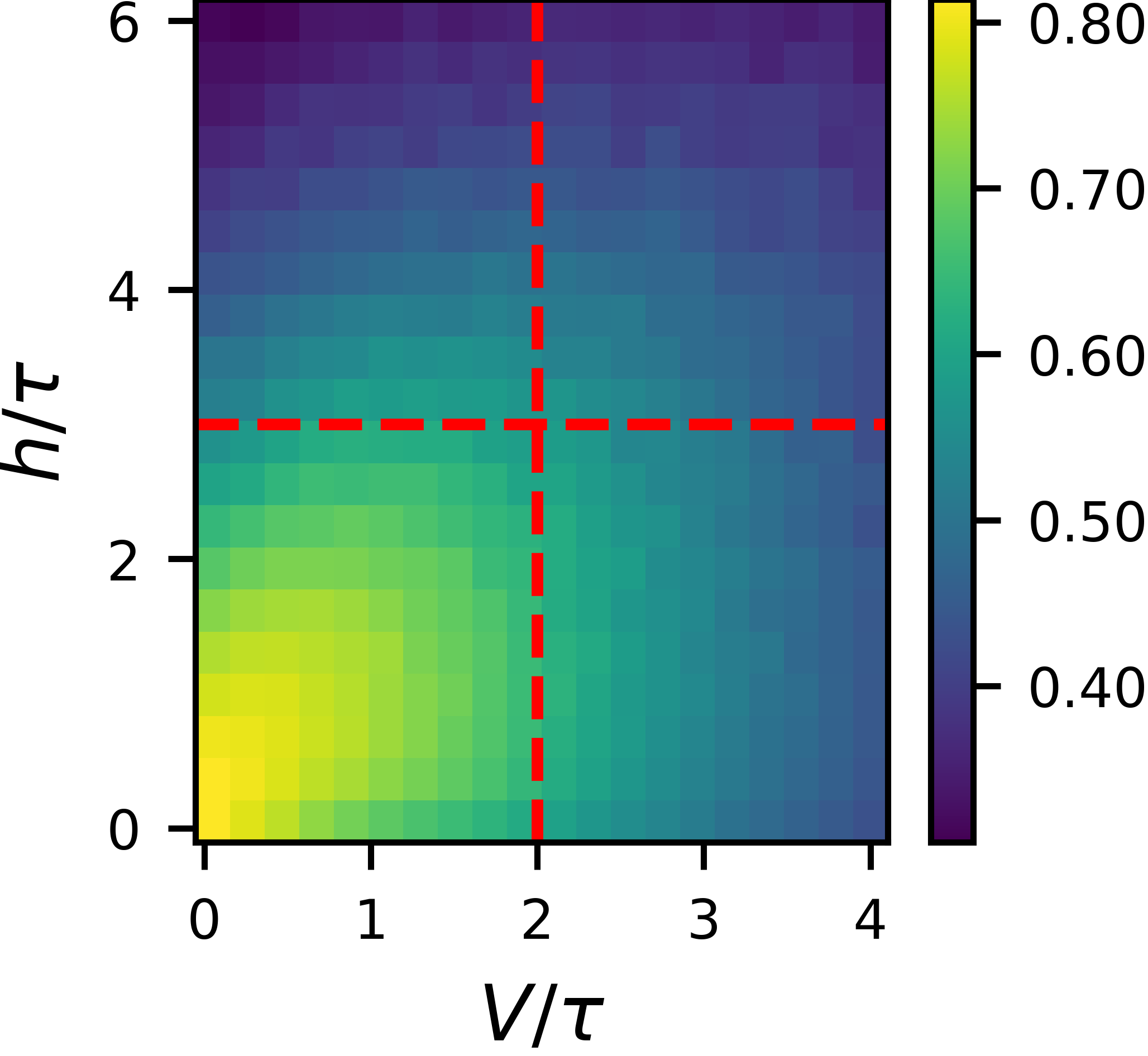}
          \caption{Bulk von Neumann Entropy}
     \end{subfigure}
     \begin{subfigure}[b]{0.3\textwidth}
          \centering
          \includegraphics[width=\linewidth]{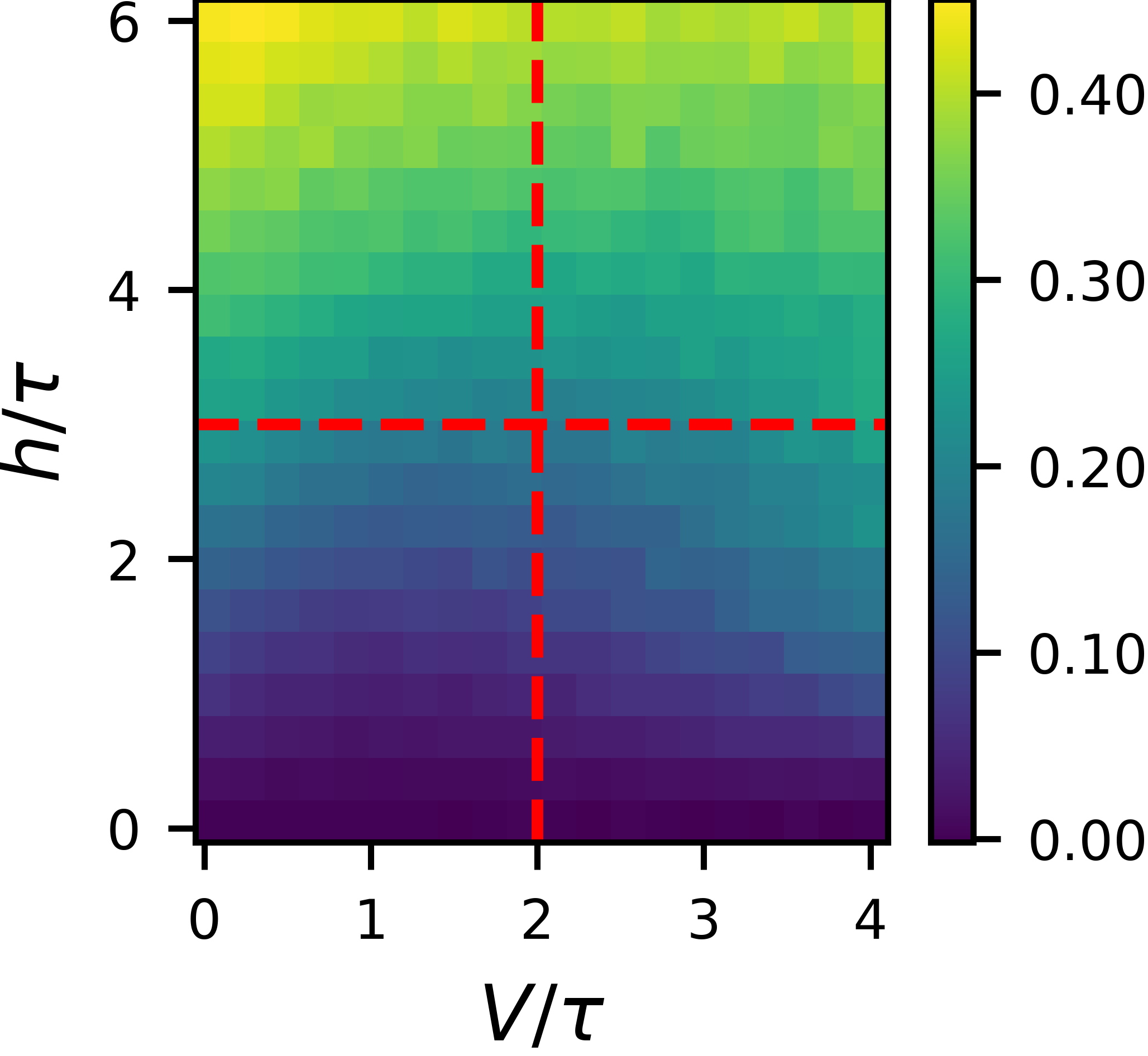}
          \caption{Bulk Imbalance}
     \end{subfigure}
     \begin{subfigure}[b]{0.3\textwidth}
          \centering
          \includegraphics[width=\linewidth]{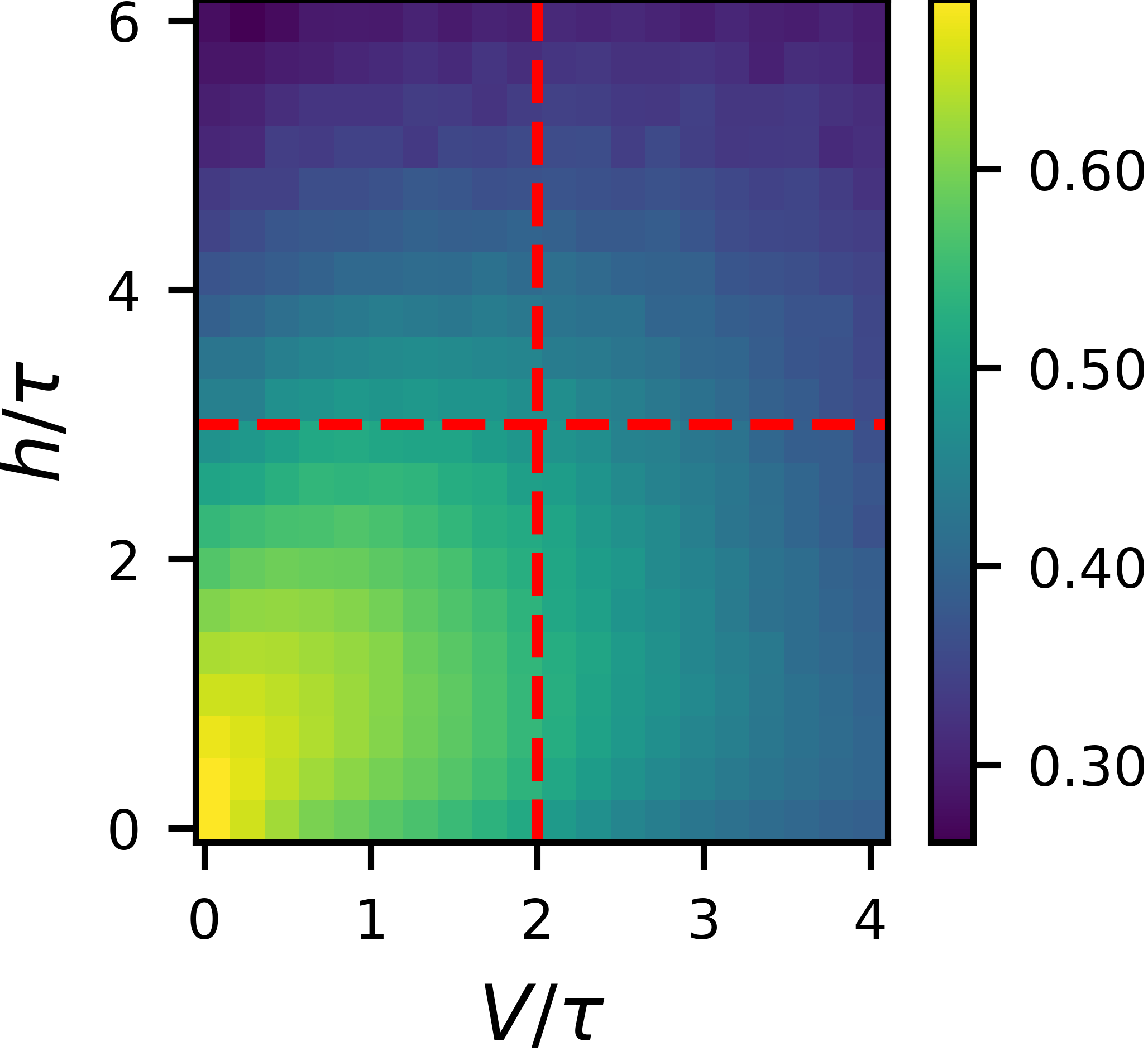}
          \caption{Bulk Number Entropy}
     \end{subfigure}
     
     \begin{subfigure}[b]{0.3\textwidth}
          \centering
          \includegraphics[width=\linewidth]{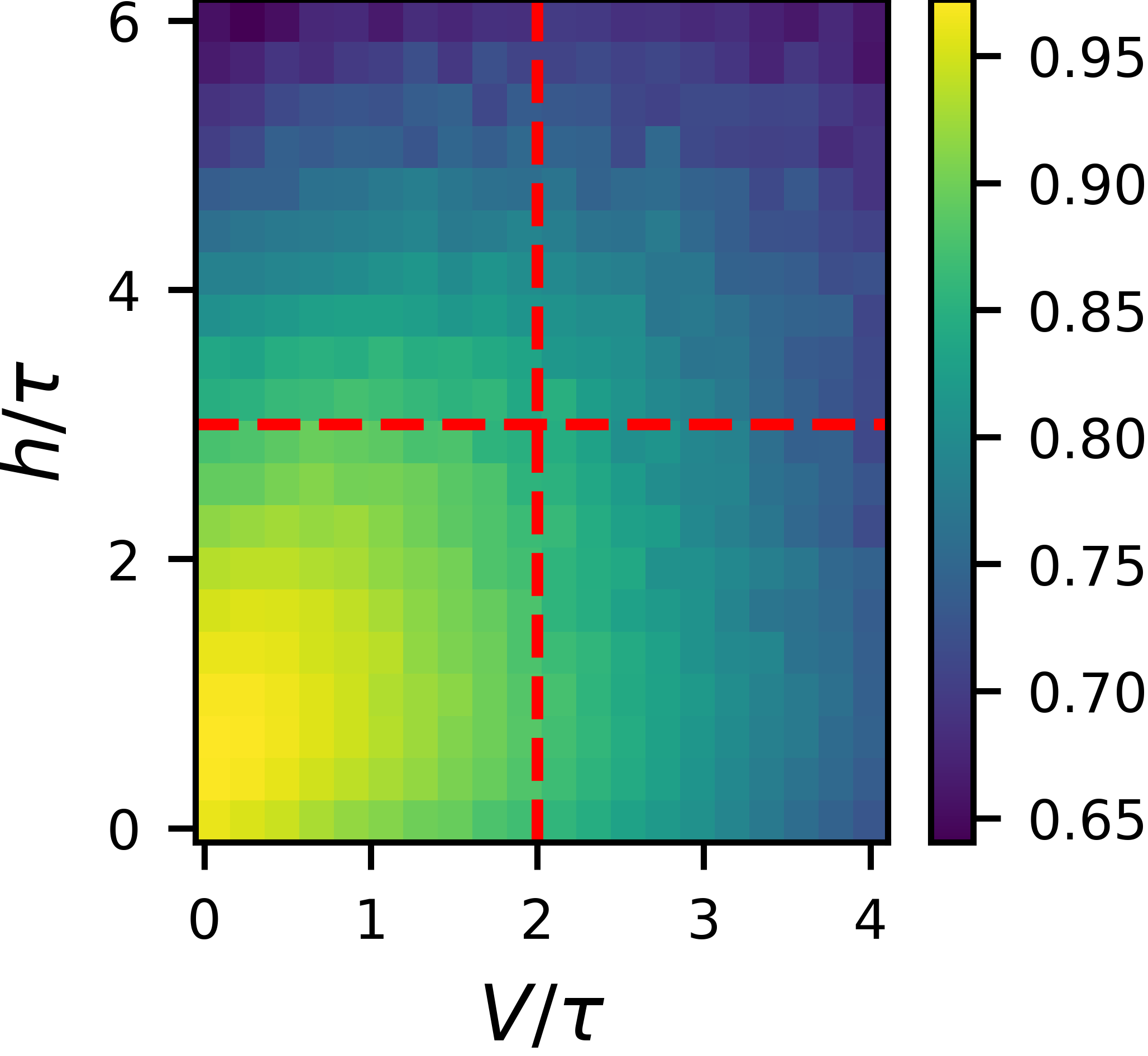}
          \caption{Local von Neumann Entropy}
     \end{subfigure}
     \begin{subfigure}[b]{0.3\textwidth}
          \centering
          \includegraphics[width=\linewidth]{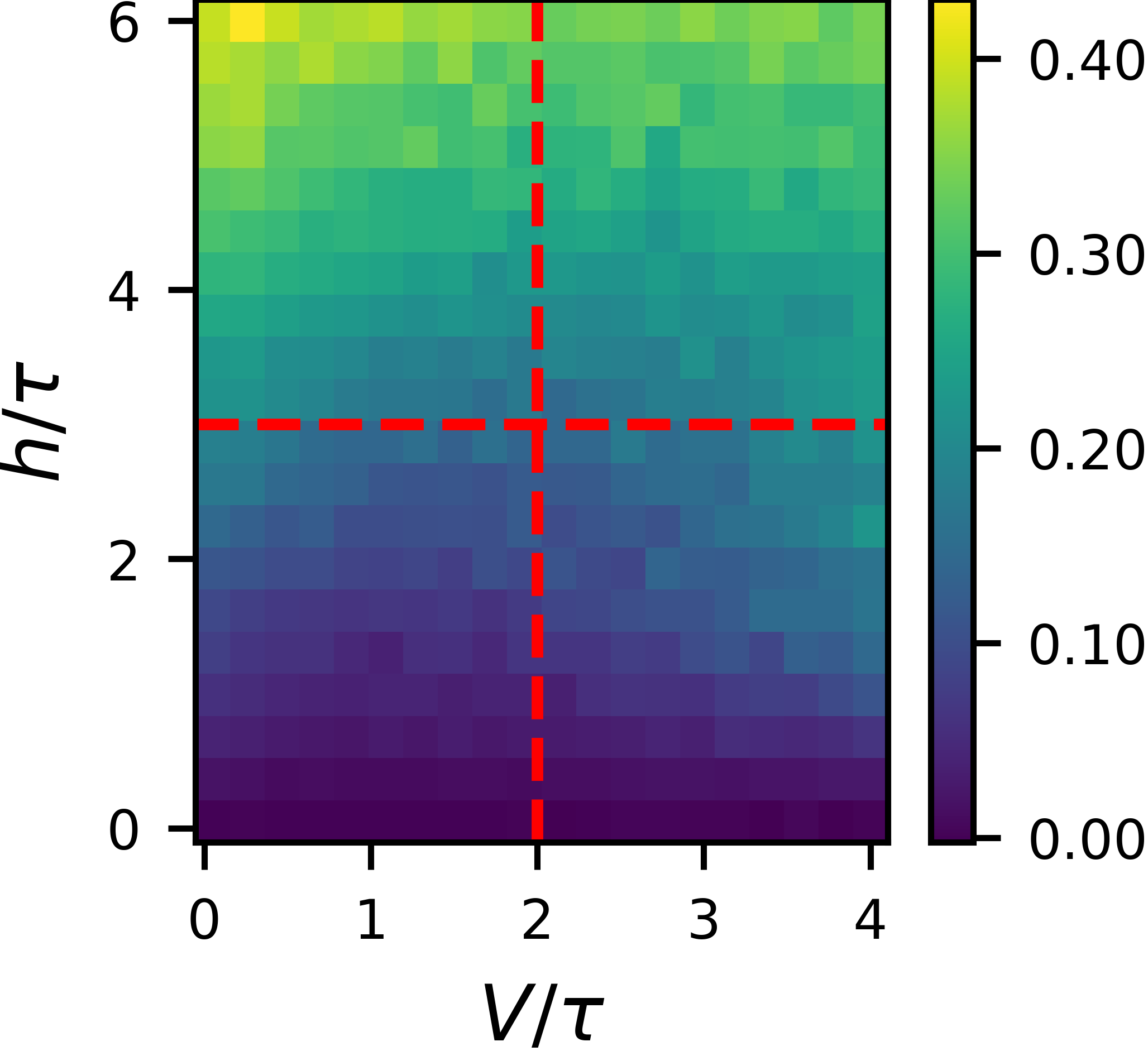}
          \caption{Local Imbalance}
     \end{subfigure}
     \begin{subfigure}[b]{0.3\textwidth}
          \centering
          \includegraphics[width=\linewidth]{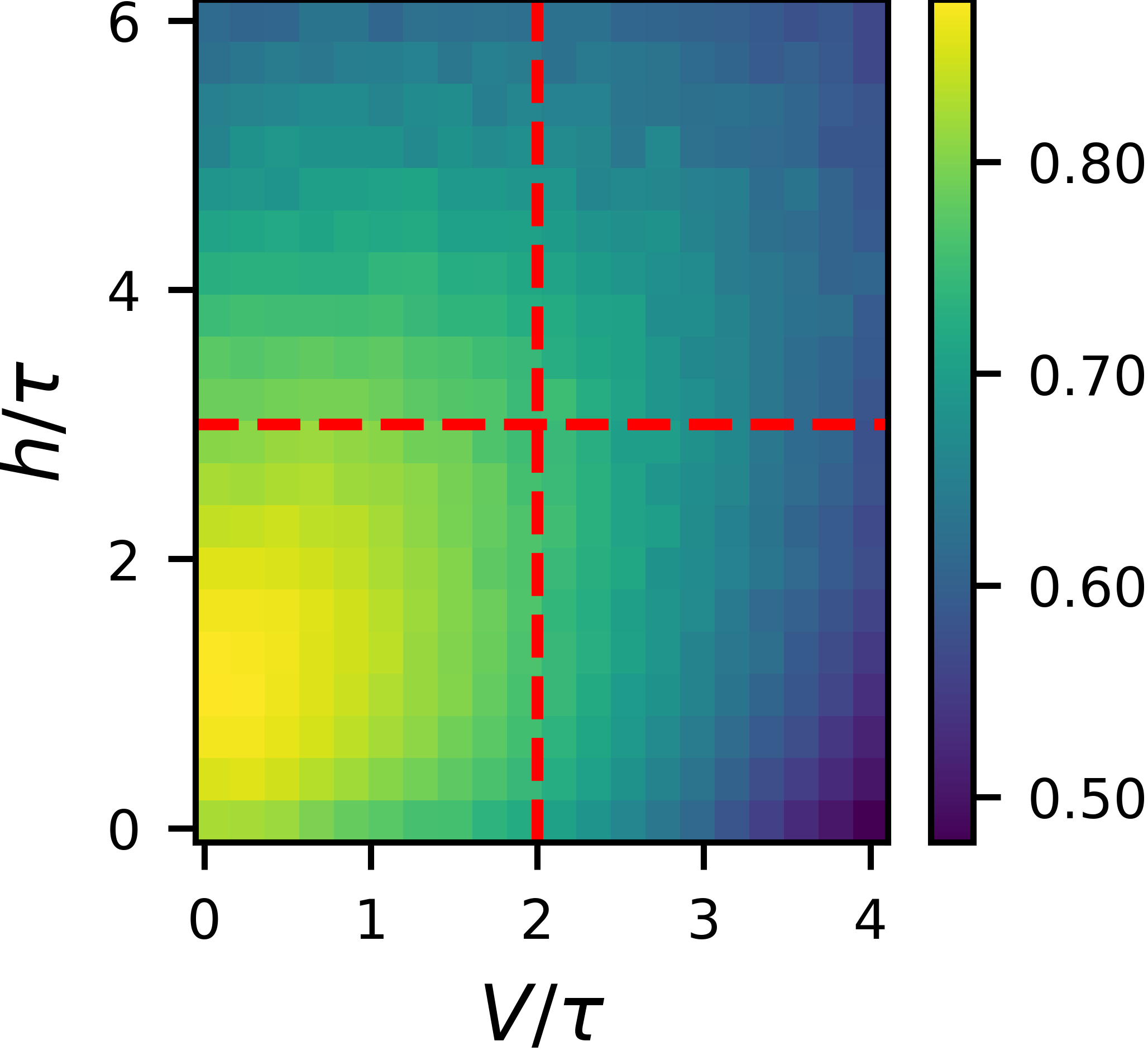}
          \caption{Local Number Entropy}
     \end{subfigure}
     \caption{Time and disorder-averaged steady-state quantities across the full $V-h$ phase diagram for the ideal system; initialized in a charge-density wave configuration. The ideal system consists of $L=10$ sites, exponential total evolution times of $\tau t_f = 10^{10}$, time-averaging over the final third of the total evolution time, and disorder-averaging over 512 samples per $(V, h)$ coordinate.} 
     \label{fig:app-phases-ideal}
\end{figure*}

\section{Ideal phase diagrams for a system of size L=10}
\label{sec:app-phases-ideal}
In \cref{sec:phases} of the main text, we analyzed the von Neumann entropy, the imbalance, and the number entropy in both bulk and local variants (see \cref{sec:sigs} of the main text for definitions and discussion) respectively. There we considered a highly conservative realistic scenario: with constraints placed on evolution times, number of disorder realizations, and parameter characterization errors. Here we present similar results but for an 'ideal' system of $L=10$ sites, exponential time scales of $\tau t_\text{f} = 10^{10}$, late-time time-averaging - wherein we average over the final third of the evolution time to eliminate remaining fluctuations - and $512$ disorder realizations. Additionally, we eliminate the absolute and relative uncertainties in the parameters $h/\tau$ and $V/\tau$ in the main text. 

\begin{figure*}
    \begin{subfigure}[b]{0.45\textwidth}
          \centering
          \includegraphics[width=\linewidth]{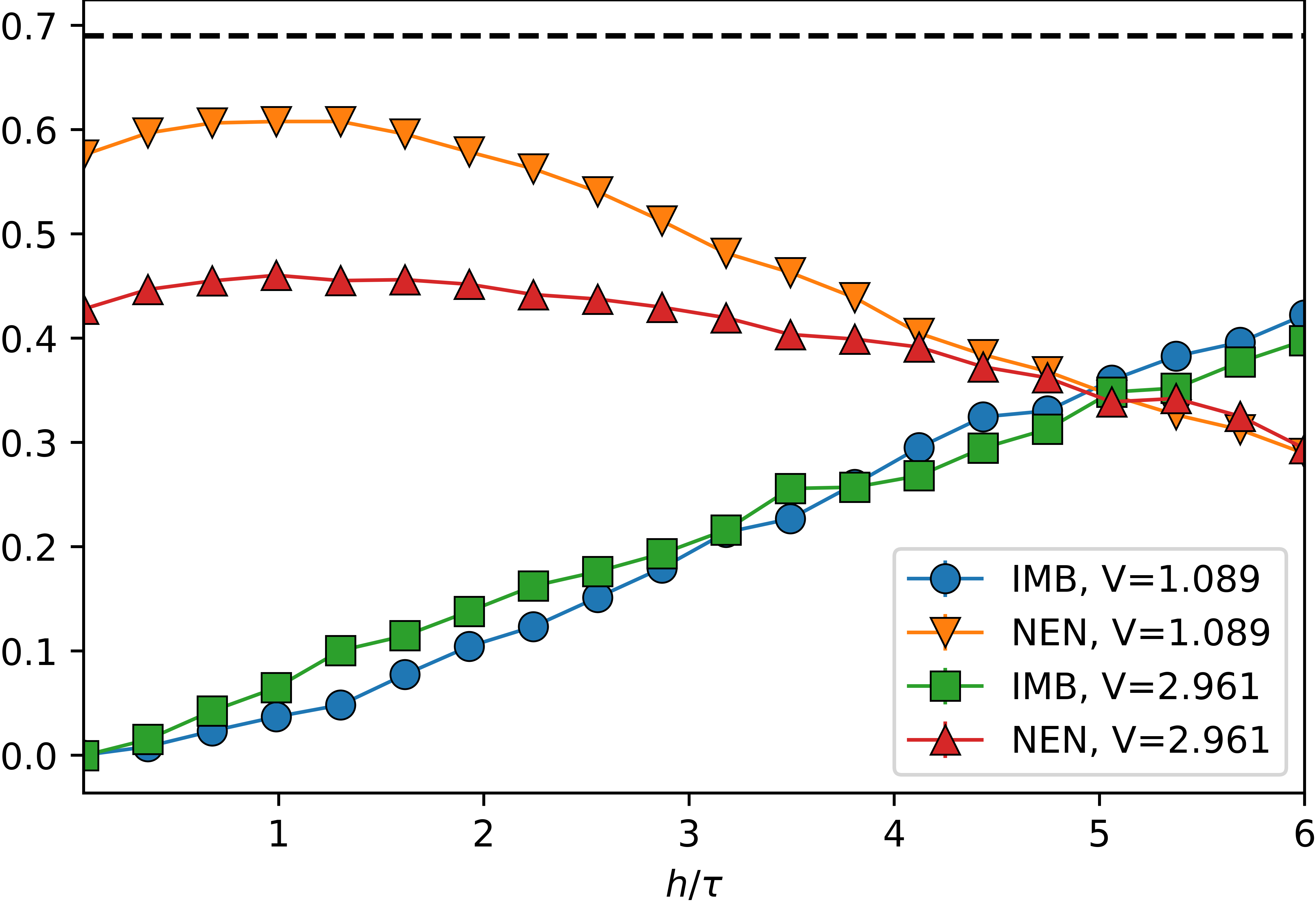}
          \caption{Bulk Imbalance and Number Entropy against $h/\tau$}
     \end{subfigure}
     \begin{subfigure}[b]{0.45\textwidth}
          \centering
          \includegraphics[width=\linewidth]{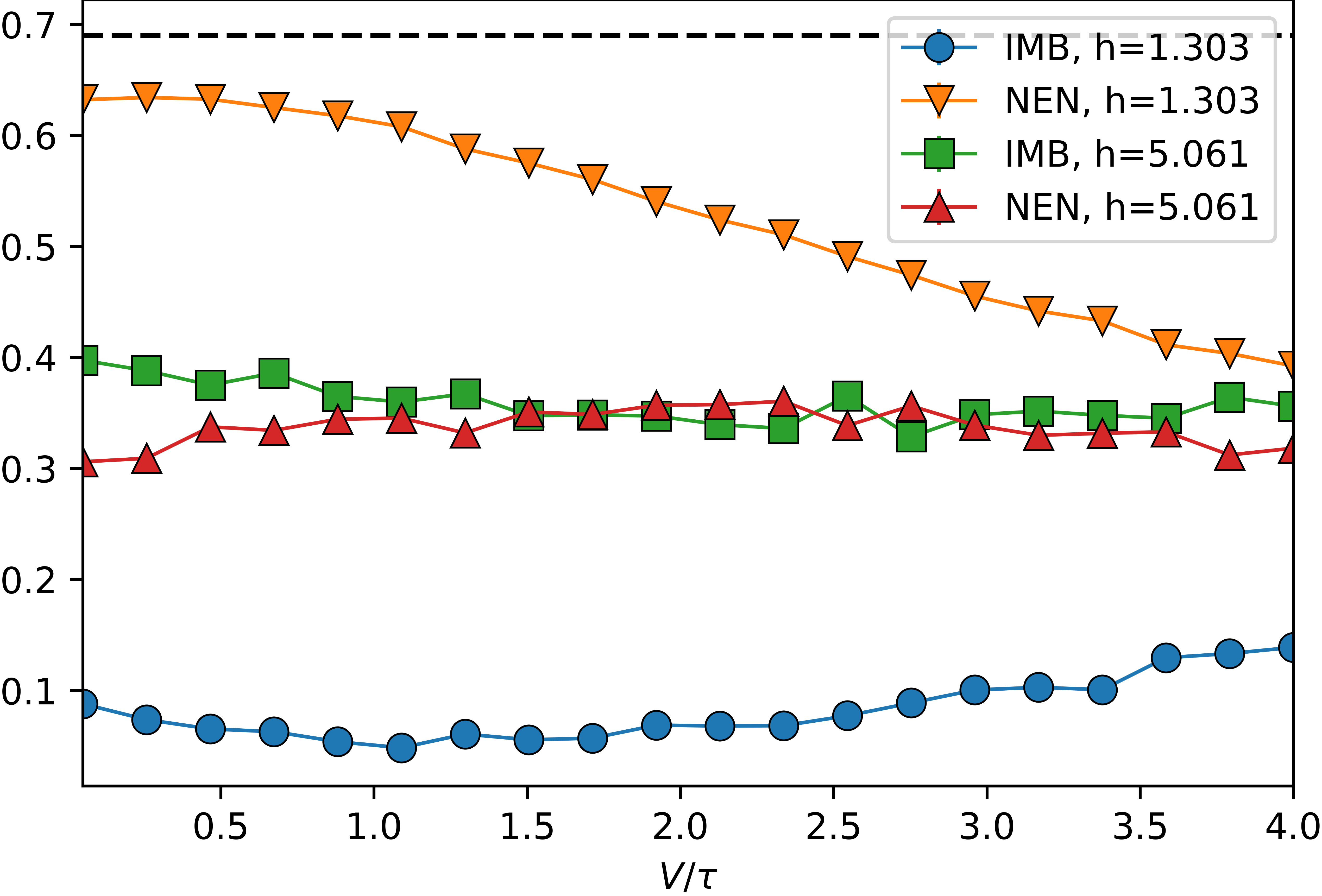}
          \caption{Bulk Imbalance and Number Entropy against $V/\tau$}
     \end{subfigure}
     
     \begin{subfigure}[b]{0.45\textwidth}
          \centering
          \includegraphics[width=\linewidth]{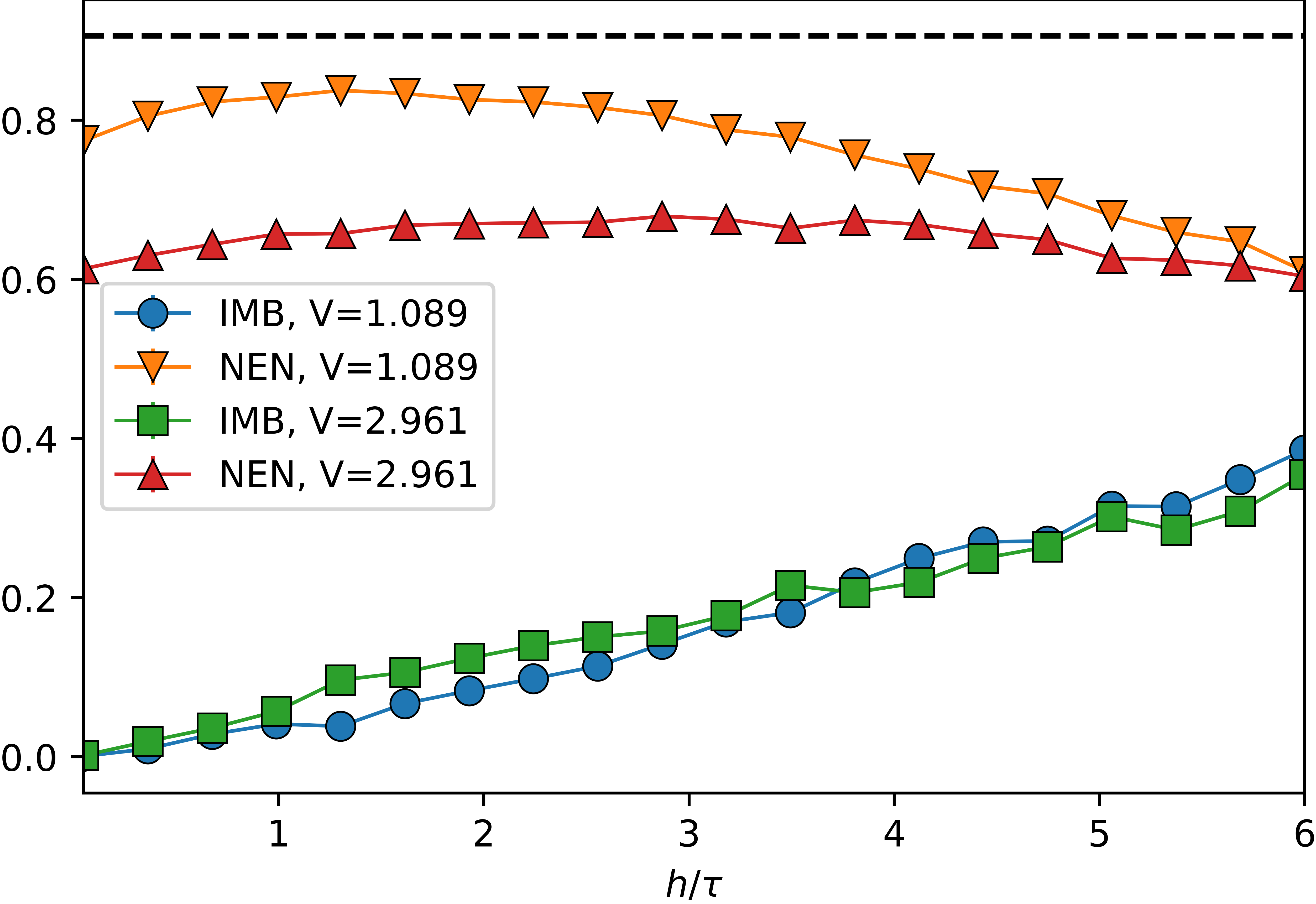}
          \caption{Local Imbalance and Number Entropy against $h/\tau$}
     \end{subfigure}
     \begin{subfigure}[b]{0.45\textwidth}
          \centering
          \includegraphics[width=\linewidth]{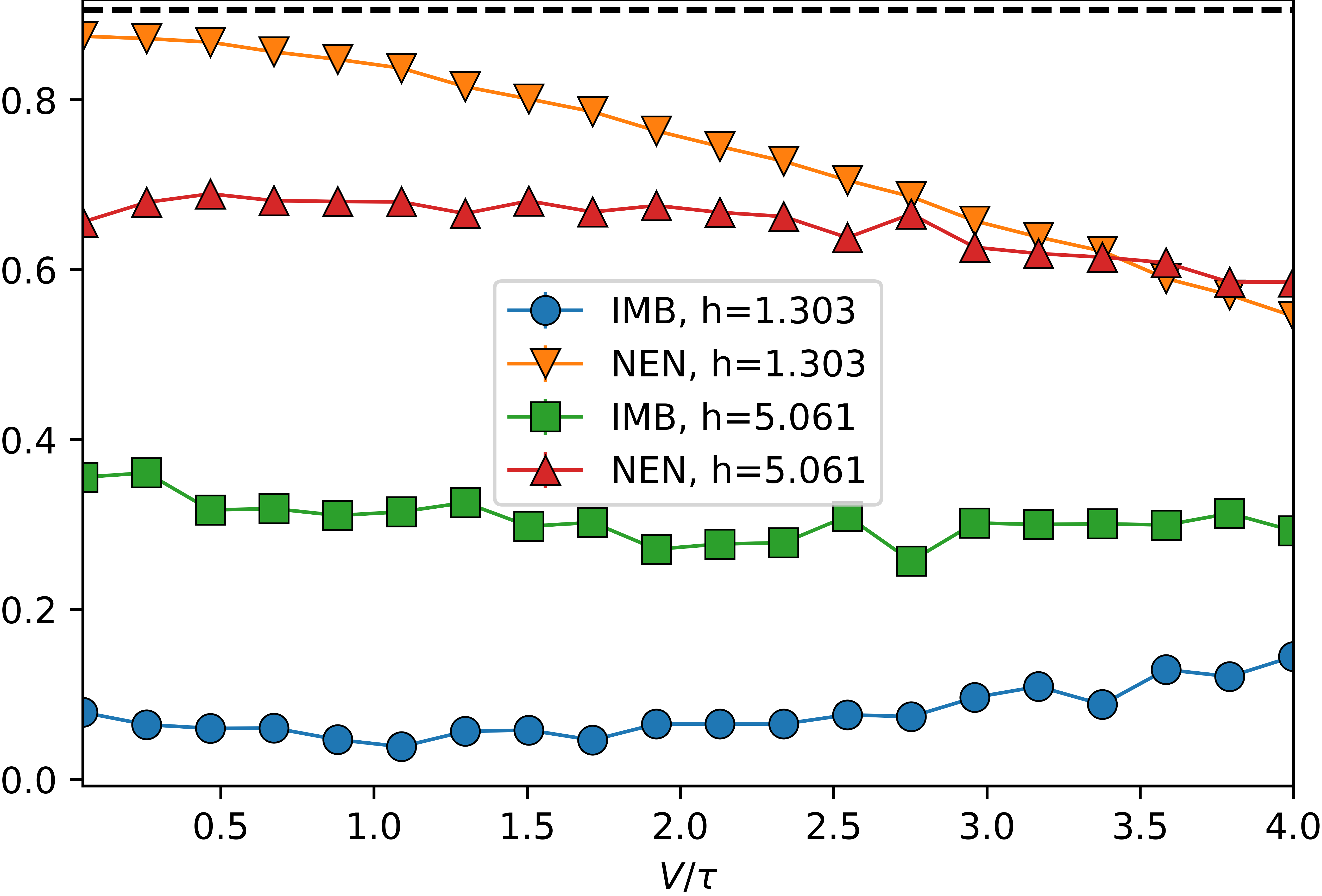}
          \caption{Local Imbalance and Number Entropy against $V/\tau$}
     \end{subfigure}
      \caption{Slices of the phase diagrams of \cref{fig:app-phases-ideal} (the ideal $L=10$ model) for \textbf{(a)} and \textbf{(c)} fixed values of $V/\tau$ on both sides of the transition point between thermal and insulating phases at $V/\tau=2$, and \textbf{(b)} and \textbf{(d)} fixed values of $h/\tau$ on both sides of the crossover from thermal to MBL regimes at  $h/\tau=3.5$; error bars shown where visible. }
      \label{fig:app-scan-ideal}
\end{figure*}
The results of this analysis of an ideal system are shown in \cref{fig:app-phases-ideal}, and are qualitatively similar to the corresponding results in the main text: the imbalance is agnostic to changes in $V/\tau$ and cannot detect the insulating regime at $V/\tau > 2$, this is in contrast to the von Neumann and number entropies which identify both this transition and the ergodic-MBL crossover at $h/\tau \sim 3.5$. Strikingly, the local variants which use measurements on only two sites in the middle of the array are qualitatively similar to the bulk variants, indicating that local measurements are enough to identify the different regimes.

Individual slices of the phase diagrams of the ideal system shown in \cref{fig:app-phases-ideal} are shown in \cref{fig:app-scan-ideal}, wherein we fix $V/\tau$ or $h/\tau$ and scan the other respectively. The behaviours noted in \cref{sec:phases} of the main text are much clearer in the larger, ideal system. The imbalance is clearly agnostic to changes in $V/\tau$, as evidenced by its constancy in panels \textbf{(b)} and \textbf{(d)} of \cref{fig:app-scan-ideal}. Moreover, the imbalance curves in panels \textbf{(a)} and \textbf{(c)} respectively are almost identical. This supports the protocol we propose which requires different post-processing on charge measurements to construct both the imbalance and number entropy in tandem. The number entropy has clearly different behaviour as both $V/\tau$ and $h/\tau$ vary, staying fixed for sufficiently high $V/\tau$ or $h/\tau$. Panel \textbf{(b)} shows this behaviour most clearly, with both the imbalance and number entropy constant with increasing $V/\tau$ for high disorder, and with imbalance constant but number entropy decreasing with increasing $V/\tau$ for low disorder.
\begin{figure}[ht]
    \centering
\includegraphics[width=\linewidth]{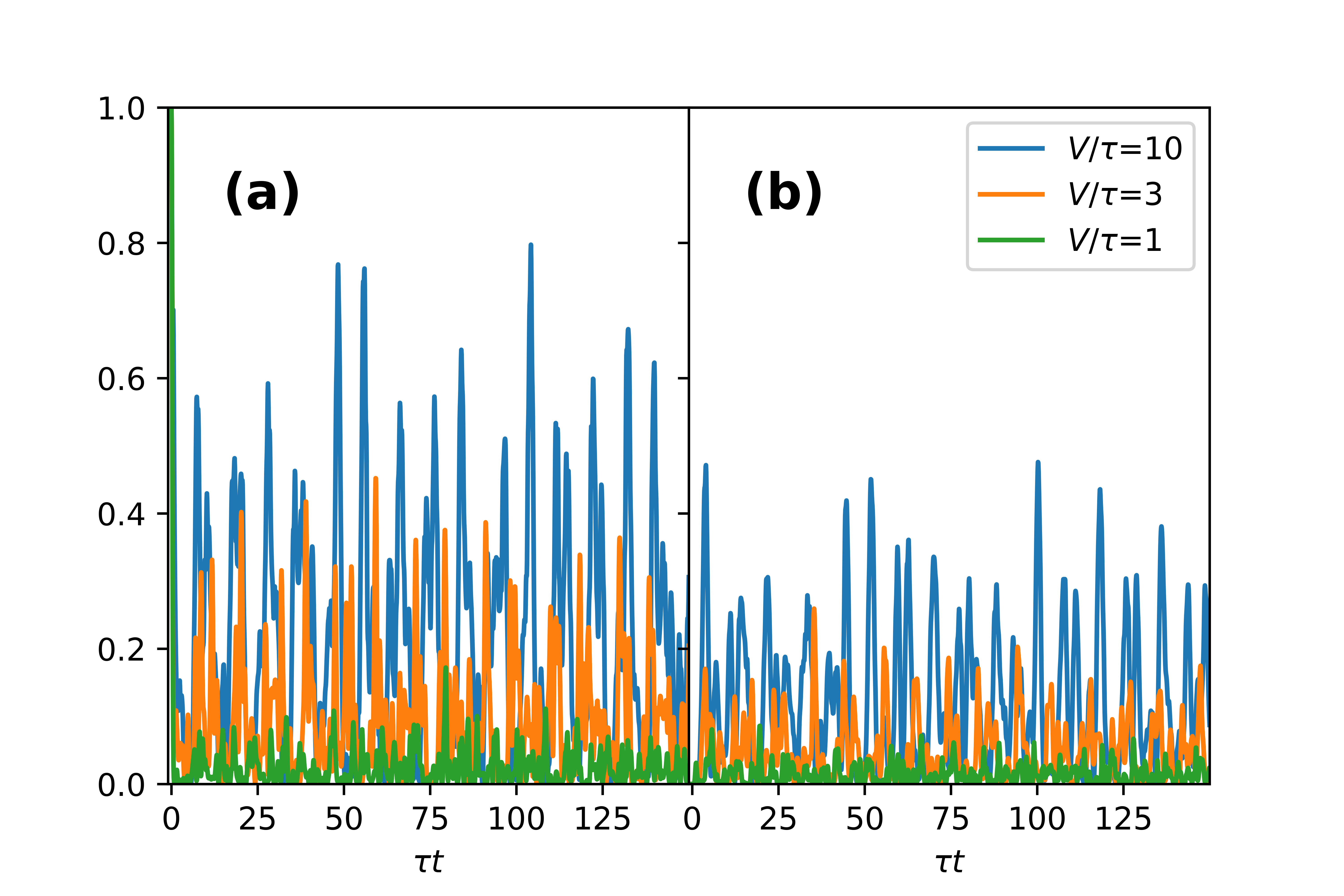}
    \caption{ Numerical evidence of coherent oscillation of a charge-density wave between two configurations as $V/\tau$. For a system initialized in the state $|\bullet, \circ, \bullet, \cdots\rangle$ \textbf{(a)} shows overlap with the initial state and \textbf{(b)} shows the overlap with the inverted state $|\circ, \bullet, \circ, \cdots\rangle$. System hamiltonian is a typical disorder-realization of strength $h/\tau=1$. }
    \label{fig:app-oscillating}
\end{figure}

\section{Oscillating Insulating State}
\label{sec:app-oscillating}

In the main text we note that a potential explanation for observing thermal values for the imbalance, but low number entropies, is by a `rolling' behaviour in which the state of the system moves coherently from left to right. Here we provide some preliminary evidence for this explanation.

We initialize a system in the charge-density wave $|\psi(0)\rangle = |\bullet, \circ, \bullet, \cdots\rangle$ (where $\circ$ corresponds to an empty dot, and $\bullet$ to an occupied dot) and select a typical disorder profile of strength $h/\tau=1$. In \cref{fig:app-oscillating} we then investigate the overlap of the state $|\psi(t)\rangle$ with $|\psi(0)\rangle$ in panel \textbf{(a)} and its overlap with the complementary charge-density wave state $|\circ, \bullet, \circ, \cdots\rangle$ in panel \textbf{(b)}. We compute the overlap as the expectation value \textit{squared}: $|\langle\psi(t)|\psi_\text{target}\rangle|^2$. For low $V/\tau$ (thermal regime), we see both curves drop rapidly to low values and remain there. But as we increase $V/\tau$ we see sharp revivals in the overlap of $|\psi(t)\rangle$ with both charge-density wave states. The first peak in panel \textbf{(a)} (after $t=0$) occurs after the first peak in panel \textbf{\textbf{(b)}} and at approximately the same time difference as the first peak in panel \textbf{\textbf{(b)}} does from $t=0$. This suggests that our `rolling' behaviour explanation may be a good approximation to the dynamics of the state; though a more detailed investigation is warranted. Such an investigation is beyond the scope of this article.

\section{Derivation of the Infinite-Temperature Thermal Number Entropy}
\label{sec:app-thermalentropy}

We start by assuming that, given $L$ sites populated by $N_0$ electrons, computational microstates which conserve $N_0$ are equally probable i.e. that the microcanonical ensemble gives the correct physical description of the equilibrated system at late time \cite{DAlessio2016}. The problem of deriving the probability distribution $p_k(n)$ of observing $n$ electrons within $k$ selected sites becomes straightforward. The probability $p_k(n=N)$ is simply the probability of detecting $N$ occupied sites and $k-N$ empty sites, multiplied by the multiplicity $k \choose N$ of such microstates:

\begin{equation}
    p_k(n = N) = {k \choose N} \underbrace{\left[\prod_{j=0}^{N-1} \frac{n_0 - j}{L - j} \right]}_\text{occupied}\underbrace{\left[\prod_{j=0}^{k-N-1} 1 - \frac{n_0 - N - j}{L - N - j} \right]}_\text{empty}
\end{equation}

The number entropy $S_N$ of such an infinite-temperature subsystem is then readily calculated according to \cref{eq:nent} of the main text with $\rho(t) \to \rho_\text{th}$. This quantity serves - in a similar capacity as the page entropy - as a thermal limiting case for the number entropy \cite{Page1993}.

\end{document}